\documentclass[twocolumn]{aastex63}
\usepackage{color}

\usepackage{gensymb}
\usepackage{amsmath}

\usepackage[flushleft]{threeparttable}
\newcommand{\unit}[1]{\ensuremath{\, \mathrm{#1}}}
\graphicspath{{./}{figures/}}
\begin{document}

%\title{Using ALMA and HST to test the IC-CMB model for high-energy emission from AP Librae}
\title{Circumnuclear Dust in AP Librae and the source of its VHE emission}

\correspondingauthor{Agniva Roychowdhury}
\email{agniva.physics@gmail.com}

\author[0000-0003-1101-8436]{Agniva Roychowdhury}
\affiliation{Department of Physics, University of Maryland Baltimore County, 1000 Hilltop Circle, Baltimore, MD 21250, USA}

\author[0000-0002-7676-9962]{Eileen T. Meyer}
\affiliation{Department of Physics, University of Maryland Baltimore County, 1000 Hilltop Circle, Baltimore, MD 21250, USA}

\author[0000-0002-2040-8666]{Markos Georganopoulos}
\affiliation{Department of Physics, University of Maryland Baltimore County, 1000 Hilltop Circle, Baltimore, MD 21250, USA}
\affiliation{NASA Goddard Space Flight Center, Code 663, Greenbelt, MD 20771, USA}

\author[0000-0003-1317-8847]{Peter Breiding}
\affiliation{Department of Physics and Astronomy, West Virginia University, P.O. Box 6315, Morgantown, WV 26506, USA}
\affiliation{Center for Gravitational Waves and Cosmology, West Virginia University, Chestnut Ridge Research Building, Morgantown, WV 26505, USA}

\author[0000-0001-6640-0179]{Maria Petropoulou}
\affiliation{Department of Physics, National and Kapodistrian University of Athens, University Campus Zografos, GR 15783, Greece}
%\collaboration{1}{(AAS Journals Data Scientists collaboration)}

%\collaboration{1}{(LaTeX collaboration)}

%\nocollaboration{2}

%% Mark off the abstract in the ``abstract'' environment.
\begin{abstract}
The broad high-energy spectral component in blazars is usually attributed to various inverse Compton scattering processes in the relativistic jet, but has not been clearly identified in most cases due to degeneracies in physical models. AP Librae, a low-synchrotron-peaking BL Lac object (LBL) detected in 2015 by H.E.S.S. at very high energies (VHE; $>$~0.5~TeV), has an extremely broad high-energy spectrum, covering $\sim$ 9 decades in energy. Standard synchrotron self-Compton models generally fail to reproduce the VHE emission, which has led to the suggestion that it might arise not from the blazar core, but on kiloparsec scales from inverse Compton scattering of cosmic microwave background (CMB) photons by a still-relativistic jet (IC/CMB). IC/CMB models for the TeV emission of AP Librae in prior works have implied a high level of infrared emission from the kpc-scale jet. With newly obtained Hubble Space Telescope imaging, we obtain a deep upper limit on the kpc-scale jet emission at 1.6 $\mu$m, well below the expected level.
High-resolution ALMA imaging in bands 3-9 reveals a residual dust disk signature after core subtraction, with a clearly thermal spectrum, and an extent ($\sim$500~pc) which matches with a non-jet residual emission seen after PSF subtraction in our 1.6 $\mu$m HST imaging. We find that the unusually broad GeV and VHE emission in AP Librae can be reproduced through the combined IC scattering of photons from the CMB and the dust disk, respectively, by electrons in both the blazar core and sub-kpc jet.
%Furthermore, since AP Librae is one in a closely interacting pair of galaxies, we believe the torus was brought about by tidal interactions and is a resulting fuel for the AGN. Both of these explanations could be further evaluated with deep and high dynamic range imaging/spectroscopy in the sub-mm and far infrared and continued monitoring of the source at TeV energies to test for variability.

%Furthermore, we also find that the VHE emission cannot be attributed to IC/CMB emission from a second electron population, in the scenario that posits a second synchrotron spectrum being responsible for the kpc-scale X-ray emission. This explanation could be tested with deep and high dynamic range imaging in the sub-mm and far infrared and/or continued monitoring of the source at TeV energies to test for variability.
%is unusually steady in the GeV band, with only occasional outbursts likely due to flares at the blazar core. We propose that the steady GeV emission is indeed attributable to IC/CMB scattering in the kpc-scale jet, but the origin of the VHE emission remains unexplained.

\end{abstract}

\keywords{galaxies: active - galaxies: jets - quasars: individual (AP Librae) - galaxies: mergers }

\section{Introduction} \label{sec:intro}

Radio-loud active galactic nuclei (AGN) are characterized by relativistic jets of fully ionized plasma moving very near the speed of light \citep[e.g.,][]{urry95,blandford2019}. High-resolution radio observations have revealed that these jets are highly collimated and originate near the central supermassive black hole \citep[SMBH, e.g., ][]{asad09}. In blazars, these jets are aligned within a few degrees to our line of sight. The emission from the fast-moving base of the jet (on parsec scales) is thus highly Doppler boosted, where the Doppler factors may range from a few to over a hundred \citep[e.g.,][]{ghis93,hovatta09}. This amplification of the emission from the base of the jet (or `core') causes what is usually termed a jet-dominated blazar spectral energy distribution (SED) from radio to gamma-rays. In a plot of log $\nu$F$_\nu$ versus log $\nu$ the typical spectrum appears as two broad components of nearly equal amplitude, the first peaking anywhere between $10^{12}-10^{17}$\,Hz and the second at GeV to TeV energies \citep[e.g.,][]{vonm95, fos98, meyer11}.

Although the cause of the lower-energy emission component in blazars is very well established as synchrotron radiation from relativistic electrons \citep[e.g.,][]{urry82}, there is less certainty regarding the location and origin of the second (higher-energy) component in blazars. In most cases it is presumed that the emission originates either from the base of the jet or very near to it ($\sim$ few parsec scales). Most theoretical models assume a purely leptonic jet by default, where the same relativistic electrons responsible for synchrotron emission inverse Compton scatter either their own synchrotron photons (synchrotron self-Compton, SSC) or photons from an external field (external Compton or EC) to produce the GeV-TeV component of the SED \citep[see e.g.,][]{bott07,sikora09}.

Hadronic models are an alternative to the purely leptonic scenario. In these cases protons are accelerated to very high energies ($\gtrsim$PeV) and can produce $\gamma$-rays directly via synchrotron radiation \citep[e.g.,][]{ahar00,muecke01, petrop15_2} or indirectly via synchrotron plus Compton processes of secondary leptons produced in photo-hadronic interactions \citep[e.g.,][]{mann91, mann93, ahar00, mucke03, petrop15}. The main objection to hadronic models is the necessarily high (often super-Eddington) power requirements \citep{sikora11,zdz15}.

%Blazars are generally characterized by their synchrotron peak frequency, which ranges over at least five orders of magnitude. While the exact dividing frequencies vary, they are usually

%While the most powerful jets are nearly all LSP blazars, low- and moderate-power jets appear in all spectral types \cite[e.g.][]{keen20}[it is 21, not 20. Also, I would not cite papers that have not been accepted].

The blazar population can be divided phenomenologically into low-synchrotron-peaking (LSP, $\nu_{peak}<10^{14}$ Hz), intermediate-synchrotron-peaking (ISP, $10^{14}<\nu_{peak}<10^{15}$ Hz) and high-synchrotron-peaking HSP sources (HSP, $\nu_{peak}>10^{15}$ Hz). Separately from the SED type, blazars with quasar-like broad emission lines (nearly all of which are LSP sources) are known as flat-spectrum radio quasars (FSRQs), suggestive of a strongly accreting system and those with an optically featureless spectrum having equivalent line widths $\Delta\lambda~<5$~\AA\, are known as BL Lacs, implying radiatively inefficient accretion. At VHE ($>$ 500 GeV), the dominant extragalactic source class is blazars and in particular, HSP BL Lacs (HBLs). While HBLs comprise $\sim$ 30\% of all identified sources at VHE only four LSP BL Lacs (LBLs) and eight FSRQs have been detected to date \citep{tevcat08, albert07, anderhub09}. In the leptonic scenario, the GeV-TeV emission is generally attributed to SSC in HBLs and EC in FSRQs and LBLs \citep[e.g.,][]{sikora09,meyer12,mades16}, possibly reflecting weak accretion disk emission in the former.

AP Librae is a well-known nearby LBL at $z=0.049$ \citep[$3.8$ kpc$/4''$,][]{jones09} and one of the only four LBLs detected at VHE \citep{hess15}. The X-ray to TeV component is unusually broad, extending over $\sim$9 orders of magnitude in frequency.  AP Librae also exhibits a resolved radio jet which has been detected in the X-rays with the \emph{Chandra} X-ray observatory \citep{cas99, kauf13}, making it the only LBL source out of approximately 200 AGN known to host an X-ray emitting kpc-scale jet\footnote{XJET archive: https://hea-www.harvard.edu/XJET/}.

The broadness of the high-energy component in AP Librae relative to the synchrotron peak makes it difficult to model using standard single-zone SSC or EC models. Even before the VHE detection of the source by H.E.S.S., \cite{tav10} noted this difficulty when attempting to fit a one-zone SSC model to \emph{Fermi}/LAT observations. Their model could not fit both the X-ray and the GeV bands simultaneously, underestimating the X-rays by $\sim$ an order of magnitude to fit the GeV or vice versa. They attributed this to the gross non-simultaneity of the Fermi-LAT and X-ray observations ($\sim$ 7 years) and the possible presence of an X-ray jet. However \cite{san15} confirmed these findings with quasi-simultaneous X-ray and \emph{Fermi}/LAT observations (see also \citealt{zac16}).

Subsequently several efforts were made to move beyond a simple SSC model for the VHE emission. Both \cite{zac16} and \cite{san15} suggested that inverse Compton scattering off cosmic microwave background photons in the extended jet \citep[the IC/CMB mechanism, first proposed by][for another source]{tav00} could explain the high-energy Fermi/LAT and H.E.S.S. emission, although the synchrotron spextrum of the large-scale jet was severely unconstrained in these works. Rather than the kpc-scale jet, \cite{hervet2015} used a more complex blob-in-a-jet emission model and showed that electrons in the pc-scale blob could reproduce the \emph{Fermi}/LAT band and VHE spectra via inverse Compton scattering of broad line region (BLR) photons and synchrotron photons from the pc-scale jet respectively. \cite{petr17} favored the core as the origin of the high-energy emission, proposing a lepto-hadronic scenario with the TeV emission primarily arising from photo-hadronic processes.

The very broad X-ray to VHE spectrum of AP Librae is unusual but not unique, and similar modeling difficulties as above have been encountered in other IBL/LBLs detected at VHE, namely BL Lacertae \citep{albert07}, S5 0716+714 \citep{anderhub09}, 3C 66A \citep{joshi07} and W Comae \citep{acciari09}. Although a single-zone SSC model has been able to explain the VHE emission in two nearby radio galaxies with a broad second component (M87; \citealt{m87} and NGC 1275; \citealt{ngc1275}) it is inadequate for all the above blazars as well as the radio galaxy Centaurus A \citep{petrop14,tanada19}; in most cases a combination of SSC and external Compton models is required.

\begin{deluxetable*}{llclllcccc}[t]
%\tablecaption{\label{table:newdata} Radio and sub-mm Observations}
\tablecolumns{6}
\tablewidth{0pt}
\label{table:vla_alma}
\tablehead{
Obs. & Band & Freq.  & Project Code  & Date       & Synthesized Beam               & LAS &  RMS               & $F_\mathrm{core}$ & $F_\mathrm{tot}$ \\
            &      &    (GHz)       &               & {\footnotesize YYYY-MM-DD} &           (arcsec)          & (arcsec)       &  (Jy)              &  (Jy)             & (Jy) }
\startdata
VLA         &  L   &   1.5         & AB700         & 1994-04-19 & $3.23\times1.17$ & 36       & $2.9\times10^{-4}$ & 1.5000 & -- \\
VLA         &  C   &   4.8         & AA099         & 1994-04-19 & $3.38\times2.84$ & 240       & $2.0\times10^{-4}$ & 1.5240 & -- \\
VLA         &  X   &   8.4         &    AV0194  & 1992-04-27 & $4.86\times2.26$ &    145   &   $5.3\times10^{-4}$      &   1.6900 & --      \\
VLA         &  U  &   15.0         &    20B-356  & 1992-04-27 & $0.24\times0.11$ &    3.6   &   $1.7\times10^{-5}$      &   2.1549 & --      \\
ALMA        &  3   &   99           & 2017.1.01411.T  & 2018-06-09          &      $4.51\times2.54$              &   24.8      &   $4.1\times10^{-5}$                 & 3.2468 & 3.2525 \\
%ALMA        &  4   &   155           & 2017.1.00107  & 2018-04-09           &         $1.25\times1.02$          &    16.2    &  $4.4\times10^{-5}$                   & 2.4845 & 2.488 \\
ALMA        &  5   &   198           & 2017.1.00568  & 2018-09-17           &        $0.35\times0.29$     &      4.8     &  $5.4\times10^{-5}$                  & 2.4037 & 2.4148\\
%ALMA        &  6   &   224           & 2016.1.00598  & 2017-08-18           &                    &   1.4      &                    & & \\
ALMA        &  6   &   232           & 2017.1.00995  & 2018-03-15           &          $0.50\times0.40$          &    5.7     & $3.5\times10^{-5}$                     & 1.9575 & 1.9630 \\
ALMA        &  7   &   337           & 2017.1.00258  & 2018-05-18          &          $0.92\times0.66$           &    7.3     & $7.1\times10^{-5}$                    & 2.3496 & 2.3550 \\
ALMA        &  7   &   341           & 2017.1.01583  & 2018-04-19           &           $0.53\times0.40$         &   5.3       &    $5.8\times10^{-5}$                & 2.0052 & 2.0073 \\
ALMA        &  7   &   348           & 2017.1.00963  & 2018-05-29           &             $0.99\times0.68$       &  7.6      &  $6.2\times10^{-5}$                  & 2.3919 & 2.3957 \\
ALMA        &  8   &   405           & 2017.A.00047  & 2018-06-30           &            $0.80\times0.57$        &   5.7      &   $1.9\times10^{-4}$                 & 2.1090 & 2.1240 \\
ALMA        &  8   &   426           & 2017.1.00239  & 2018-06-01           &          $0.79\times0.65$          &  6.0       &  $2.9\times10^{-4}$                  & 2.0920 & 2.0990 \\
ALMA        &  8   &   464           & 2013.1.00244  & 2015-05-20          &             $0.37\times0.27$       &   2.7      & $2.4\times10^{-4}$                   & 1.0390 & 1.0450 \\
ALMA        &  9   &   654           & 2017.1.01555  & 2018-05-20           &          $0.50\times0.36$          &  3.8       &  $4.8\times10^{-4}$                     & 1.9420 & 1.9536 \\
ALMA        &  9   &   671           & 2017.A.00047  & 2018-06-30           &           $0.42\times0.36$         &  3.4       & $9.4\times10^{-4}$                    & 1.7510 & 1.7670 \\
ALMA        &  9   &   679           & 2017.1.00337  & 2018-05-23           &          $0.44\times0.34$          &  3.7       &    $6.8\times10^{-4}$                & 1.9240 & 1.9353 \\
ALMA        &  9   &   692           & 2017.1.00023  & 2018-05-04           &           $0.36\times0.21$         &   2.7      &  $1.0\times10^{-3}$                  & 1.8453 & 1.8665 \\
ALMA        &  9   &   699           & 2017.1.01555  &  2018-05-20          &           $0.47\times0.34$         &  3.5       &   $9.9\times10^{-4}$                 & 1.9070 & 1.9200 \\
\enddata
\caption{Radio and sub-mm Observations. LAS: Largest Angular Scale; RMS: root mean squared sensitivity.}
\end{deluxetable*}

For AP Librae, the potential physical scenarios and emission mechanisms laid out in previous works are considerably at variance, and imply very different jet characteristics at a fundamental level, including jet composition, velocity, and total energy content.
%A further discussion of possible observational discriminants is deferred to the discussion in this paper.
Our aim in this study is to explain the very broad second component in AP Librae and in particular to test the clear predictions of the models invoking IC/CMB in the kpc-scale jet for the TeV emission \citep[e.g.][]{zac16}. These models predict a high level of infrared synchrotron emission from the jet, provided by the same electrons required to explain the TeV emission. In this paper, we present new observations of AP Librae with the Very Large Array (VLA), Atacama Large Millimeter/sub-millimeter Array (ALMA), the Hubble Space Telescope (HST), and the \emph{Fermi}/LAT observatory designed to fully constrain the synchrotron spectrum of the kpc-scale jet and deduce the origin of the VHE emission.

In Section \ref{sec:observations}, we present the multi-wavelength data reduction procedure. In Section \ref{sec:3}, we use multi-wavelength SED modelling to discuss the origin of TeV emission from the source. In Section \ref{sec:conc}, we conclude with a summary of the work and possible future directions this study can take.

%%%%%%% OBSERVATIONS %%%%%%%%%%%%%%%%%%%%%%%

\section{Observations}
\label{sec:observations}

We describe here the VLA, ALMA, HST, and \emph{Fermi}/LAT data analyzed for this project, where we are primarily focused on new observations of the large-scale jet. Core fluxes (or total fluxes, presumed to be dominated by the core) were taken from the NASA Extragalactic Database\footnote{https://ned.ipac.caltech.edu/} unless otherwise noted.
In addition, we have utilized other observations reported previously in the literature for this source. These include \textit{Chandra} X-ray fluxes of the large-scale jet which we obtained from \cite{kauf13}, and the TeV spectrum and flux points from the H.E.S.S. observatory \citep{hess15}.

\subsection{Very Large Array}
\label{sec:vla}
We reduced several historical VLA observations of AP Librae using the Common Astronomy Software Applications \citep[CASA;][]{casa}. A summary of these observations with the relevant image properties including peak and total fluxes, final image RMS in Jy, synthesized beam size in arcsec and largest angular scale are listed in Table \ref{table:vla_alma}. We applied standard initial calibration to all the datasets. AP Librae is core-dominated and very bright in all radio imaging and thus phase and amplitude self-calibration were applied to improve the final imaging, where we used Briggs weighting with robustness=0.5 \citep{briggs95}.

\begin{figure*}
  \includegraphics[width=\linewidth]{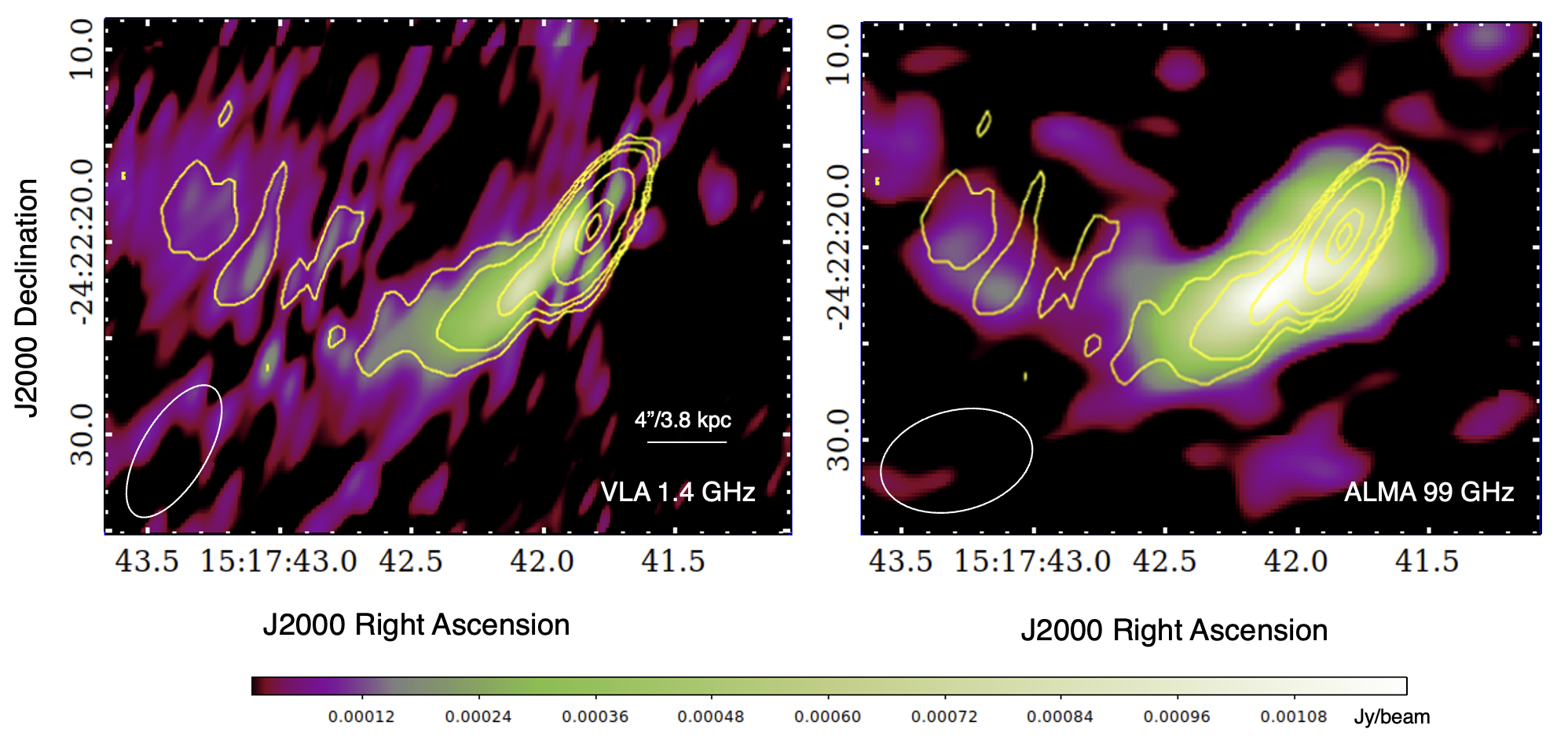}
  \caption{At left, a residual VLA L-band image of AP Librae after core subtraction. Overlaid contour lines from the same image (not core-subtracted) are drawn at 0.1, 0.2, 0.5, 5, 60, and 1000 mJy/beam. At right, a residual ALMA Band~3 image after core subtraction is shown with the same L-band contours overlaid (note that this image is shown for illustration and was not used for measuring the jet/core flux: see text). The flux scales for the images are not identical but within 10\% of that shown in the scalebar. The beam sizes are also shown at the bottom left corner of both the figures.}
  \label{fig:aplib_radio}
\end{figure*}

Previous radio imaging shows AP Librae to host a $\sim$20$''$ long radio jet \citep{cas99}. Our imaging deconvolution showed components in the large-scale jet within 1$''$ from the peak core position. Accordingly we produced core-subtracted images to more accurately measure the total jet flux. To do so we used the \texttt{clean} deconvolution task in CASA to create a point source model for only the core and then subtracted it from the total visibility data using CASA task \texttt{uvsub}. This was followed by a final \texttt{clean} of the hence core-subtracted visibility to produce a final image without the core. The total jet flux density for each of the VLA images was then given by the flux density of a large region containing all of the extended jet emission. The L-band (A-config) VLA image of AP Librae (core-subtracted) is shown at left in Figure~\ref{fig:aplib_radio}, with contours from the same observation overlaid.  %\textcolor{blue}{We have labeled the most prominent knot in the jet as knot ``A''.}
An ALMA band 3 image (also core-subtracted) on the same scale and with the same L-band contours (further described below) is shown at right.

\begin{deluxetable}{cccccc}[t]
\tablecaption{\label{table:newdata} Radio and sub-mm fluxes for AP Librae after core subtraction.}
\tablecolumns{6}
\tablewidth{0pt}
\label{table:alma}
\tablehead{
Central Freq.  & $F_{jet}$ & $F_\mathrm{jet+dust}$ & Std. Dev. \\
           (GHz)       &                (Jy)             & (Jy) }
\startdata
1.52\tablenotemark{$\dagger$} & 0.0940 & --  & 0.0010 \\
4.86\tablenotemark{$\dagger$}& 0.0450  & -- & 0.0045 \\
8.4\tablenotemark{$\dagger$} & 0.0300  & -- & 0.0030 \\
15\tablenotemark{$\dagger$} & 0.0187  & -- & 0.0037 \\
99           & -- & 0.0057 & 0.0006\\
198          & -- & 0.0070 & 0.0006\\
232          & -- & 0.0055 & 0.0004 \\
337          & -- & 0.0033 & 0.0007\\
341          & -- & 0.0027 & 0.0005 \\
348          & -- & 0.0025 & 0.0007\\
405          & -- & 0.0082 & 0.0013 \\
426          & -- & 0.0020 & 0.0008\\
464          & -- & 0.0036 & 0.0026 \\
654          & -- & 0.0106 & 0.0023 \\
671          & -- & 0.0160 & 0.0032 \\
679          & -- & 0.0115 & 0.0025 \\
692          & -- & 0.0212 & 0.0065 \\
699          & -- & 0.0110 & 0.0052 \\
\enddata
\begin{tablenotes}
\item{$^\dagger$} For each of the four VLA/JVLA observations, the flux reported here is entirely due to the jet.
\end{tablenotes}
\end{deluxetable}

\subsection{Atacama Large Millimeter/sub-millimeter Array}
\label{sec:alma}
Because AP Librae has a bright compact core in the sub-mm, it is frequently used as a calibrator source.  We analyzed a number of archival ALMA observations taken in bands 3$-$9 to better constrain the synchrotron spectrum of the kpc-scale jet in AP Librae. Details of the observations are listed in Table~\ref{table:vla_alma}, where we give the project code, central frequency, beam size, final image RMS and core and total flux. In all cases we used the appropriate CASA pipeline version to calibrate the data and prepare the measurement set (MS) for imaging with \texttt{clean}.  We used several rounds of (non-cumulative) phase-only self-calibration and a final amplitude and phase self-calibration to greatly improve the dynamic range and sensitivity of the final images.

\begin{figure*}[!t]
\centering
  \includegraphics[width=0.85\linewidth]{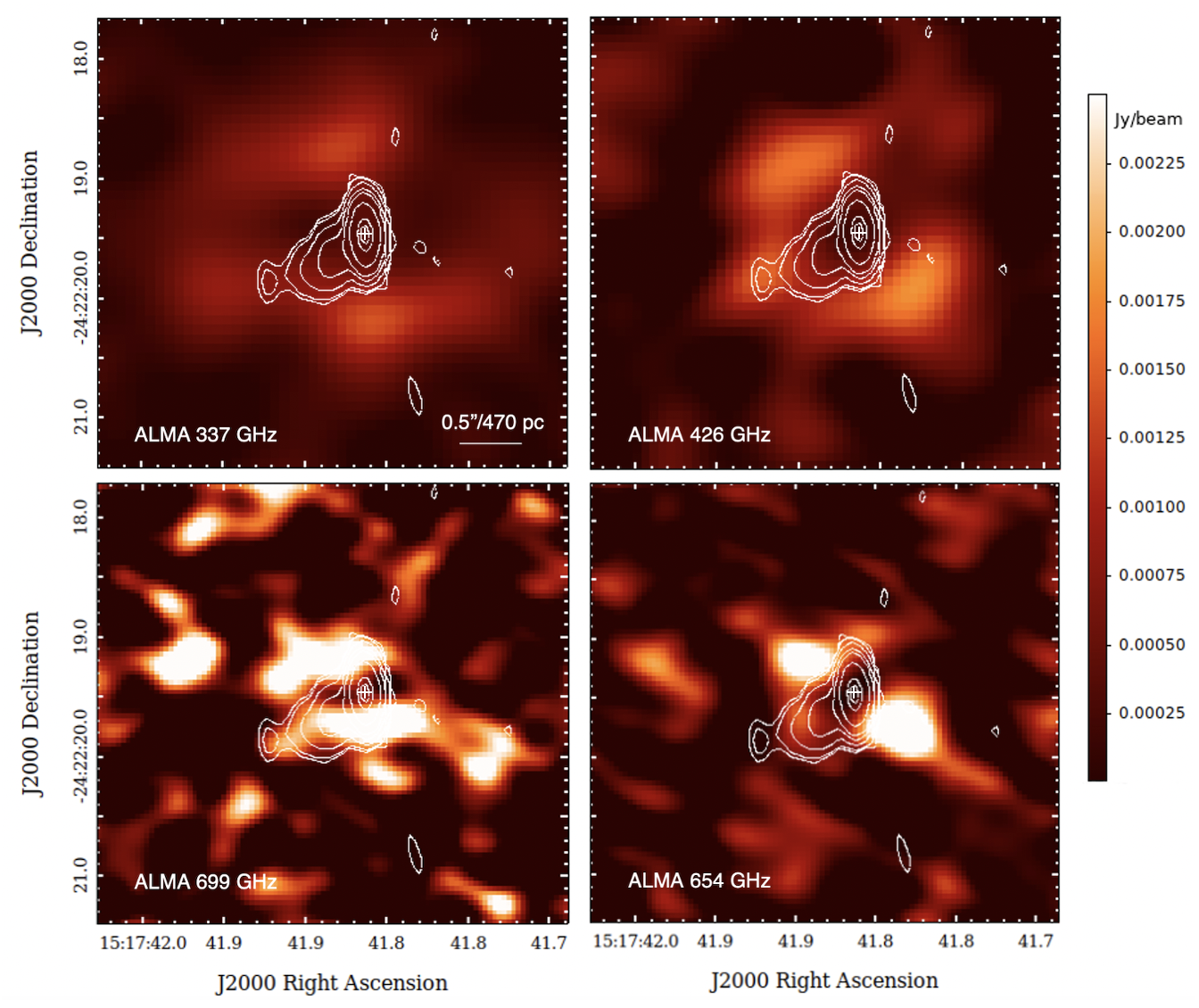}
  \caption{Core-subtracted ALMA images across a range of frequencies, with corresponding VLA 15 GHz contours superimposed. The location of the core is marked with a white cross and symmetric emission on its either side clearly increases with increase in observing frequency. It is to be noted that this quasi-thermal emission is, in actuality, continuous (Figure \ref{fig:aplib_radio}) and this symmetric effect is only due to over-subtraction of the central region which can be remedied by reducing the number of iterations per \texttt{clean} cycle. The procedure has been discussed further in Appendix \ref{sec:app1}.}
  \label{fig:alma}
\end{figure*}

We show at right in Figure \ref{fig:aplib_radio} a band 3 image with L-band radio contours overlaid. In all of the ALMA images, the point-source core is significantly ($\sim$100 times) brighter than the jet. The residual images after core-subtraction show both emission from the jet, as well as residual flux around the core, which is increasingly dominant with increase in frequency. The residual emission region appears elongated in the NE-SW direction at position angle $\sim 50$ degrees with a maximum extent of $\sim$1$''$ (about 900 pc). This can be seen in the residual core-subtracted images shown with the same flux scale in Figure \ref{fig:alma} where this emission clearly increases with increase in observing frequency. The dominance of this residual emission at higher frequencies and on similar scales to the inner jet made it difficult to accurately measure the flux from the jet alone. This was compounded by our incomplete knowledge of its structure, and the possibility of `over-subtracting' the core. To avoid the latter issue, we assume the peak flux is the total core emission, and subtract this from the total source flux, measured using a large region around the full source and extended jet. The result is a combination of the jet and residual emission. This is given along with a 1$\sigma$ error  in Table \ref{table:alma}. The associated uncertainties of the method have been described in more detail in Appendix \ref{sec:app1}. Summarizing, all the images in Figures \ref{fig:aplib_radio} and \ref{fig:alma} have been produced by subtracting the \texttt{clean} model for the core, but we followed the above method for calculating the jet+residual flux for the ALMA images only.
%At 654 GHz, the detection significance is $5\sigma$.

The residual emission in Figure \ref{fig:alma} appears as a slightly asymmetric disk. Subtraction of the core in \texttt{clean} with lower iterations per cycle would show a more continuous disk emission, like in Appendix \ref{sec:app1}. The disk has a larger NE-SW extent, while the SE harbors the blazar jet and the NW is faint. This fits with the expectation that the jet be roughly perpendicular to the disk. This has been found for FRI radio galaxies with circumnuclear dust \citep[e.g.,][]{dekoff00,landt10,drouart12}, and thus may be expected for their possible beamed counterparts, which are BL Lac objects. The residual disk emission follows a quasi Rayleigh-Jeans spectrum (Figure \ref{fig:sed_zach} and Section \ref{sec:dust}) and is therefore most likely dust emission. Since AP Librae is a blazar, the jet presumably forms a very close angle to the line of sight. If the disk axis is perfectly aligned with that of the jet, we should see a more face-on view of the dusty disk as a circular halo after core-subtraction. However, we see an absence of emission in the direction of the jet and the counter-jet. It is possible that the dusty disk is slightly misaligned with the direction of the jet, resulting in the jet blowing away a fraction of the disk in the SE-NW direction, thereby resulting in a ``gap" where the dust brightness is low. This is also consistent with X-ray observations of the core of AP Librae where \cite{kauf13} find a hydrogen column density $N_H\sim10^{21}\unit{cm^{-2}}$, a value typical of ordinary galactic absorption instead of obscuring tori due to AGN or starburst heating \citep{hickox18}.

%If a jet undergoes a strong interaction with the environment, this may manifest itself as a stationary shock which will energize the jet. Indeed, the brightest jet knot in the ALMA image occurs very close to $\sim0.5$ kpc in the SE direction, coincident with the assumed location of the torus.

%The orientation of this disk  that is being viewed edge-on.
To minimize biases due to resolution, we calculate the true size of the excess emission by deconvolving it from the beam. The dimensions could be determined from a simple Gaussian fit to the 652 GHz image inside CASA, for which the deconvolved major and minor axes are $1.30\pm0.05$ kpc and $0.20\pm0.02$ kpc. These translate to $R=0.65\pm0.02$ kpc and $W=0.20\pm0.02$ kpc where $R$ and $W$ are the NE-SW extent and the spread of the disk across the plane of the sky respectively. However, the intrinsic scale height $H$ of the disk cannot be determined because of its $\sim$ face-on orientation. So we assume $H\sim R/5$ based on the ALMA observations of CO lines in radio-quiet AGN \citep[e.g.,][]{honig19}.
%Secondly, the high-resolution ALMA observation is blind to emission larger than the LAS. From Table \ref{table:alma}, the high-frequency/resolution observations have typical LAS $\sim2-4''$, or linear size $\sim2-4$ kpc. Since we observe a size $\sim4$ kpc in the 99 GHz image (Figure \ref{fig:aplib_radio}) and a size $\lesssim2$ kpc in the high-frequency images in Figure \ref{fig:alma}, it is clear no emission has been ``resolved out" and we have determined the intrinsic size of the excess emission.
%According to the ALMA spectrum, it represents thermal emission from a large-scale molecular torus or circumnuclear disk, detected for the first time in a BL Lac. This is in marked contrast with studies that have otherwise failed to detect any dust emission from a torus in a large sample of BL Lacs (Plotkin et al. 2012).

\subsection{Hubble Space Telescope}
AP Librae has been imaged twice by HST since the SM1 optics correction. The initial observation was taken as part of a BL Lac snapshot program (6363) in 1997, consisting of 3 100 second exposures with WFPC2/PC in the F702W filter. The upper limit on the emission from the bright inner jet from this shallow observation is not constraining on the IC/CMB model predictions. We thus obtained a much deeper (2 orbit) observation with the 4th-generation Wide-field Camera 3 (WFC3) UVIS/IR imager on HST. Observations totaling 5016 seconds were taken on 15 February 2018 with the F160W filter which peaks at 1.5 microns.

In order to more clearly observe any possible jet-related emission in our near-IR imaging, it was necessary to model and subtract both the host galaxy emission and the central point-source due to the bright blazar core. For our PSF model we used an observation of a bright star made in the same instrument/filter setup and close in detector coordinates to our observation of AP Librae. We chose the bright star \texttt{V* AY Ori (05:36:08.3 -06:48:36.36)} observed in project \texttt{14695-ID7Z03010} on 14 September 2016 for 2.4 ks. The star has clear diffraction spikes of a similar strength to the central point source of AP Librae without adjacent contamination from other bright objects.

\begin{table}[!t]
\begin{center}
\caption{\label{table:galfit}Average best-fit parameters for GALFIT modeling of AP Librae. Further details of GALFIT modeling can be found in \cite{peng02}. Here, IM refers to Integrated Magnitude, $R_e$ is the half-light radius in a S\'{e}rsic profile, $R_s$ is the e-folding length in an exponential disk (expdisk), $n$ is the S\'{e}rsic index and PA is the position angle of the profile which measures an effective tilt from the vertical.}
{\centering
\begin{tabular}{cccc}\\
\tableline\tableline
Parameter &  PSF & S\'{e}rsic & Expdisk   \\
\tableline
IM & 16.6965 & 16.7258 & 19.7040\\
$R_e$, $R_s$ (pixels) & -- & 43.6671 & 6.0495 \\
$n$ & -- & 4.0540 & -- \\
Axis ratio & -- & 0.8736 & 0.9021 \\
PA & -- & 2.3710 & -55.8114 \\
Sky bckg (count rate) & 0.9622 & 0.9622 & 0.9622  \\
\tableline
$\chi^2$/D.O.F & 4.9
\\
\tableline
%2-8 keV		&	10$^{-4.4}$	&	-1.17 $\pm$ 0.01\\
%a & b & c & d & e & f
\tableline
\end{tabular}
}
\end{center}
\end{table}

\begin{figure*}[!t]
\centering
  \includegraphics[width=0.7\linewidth]{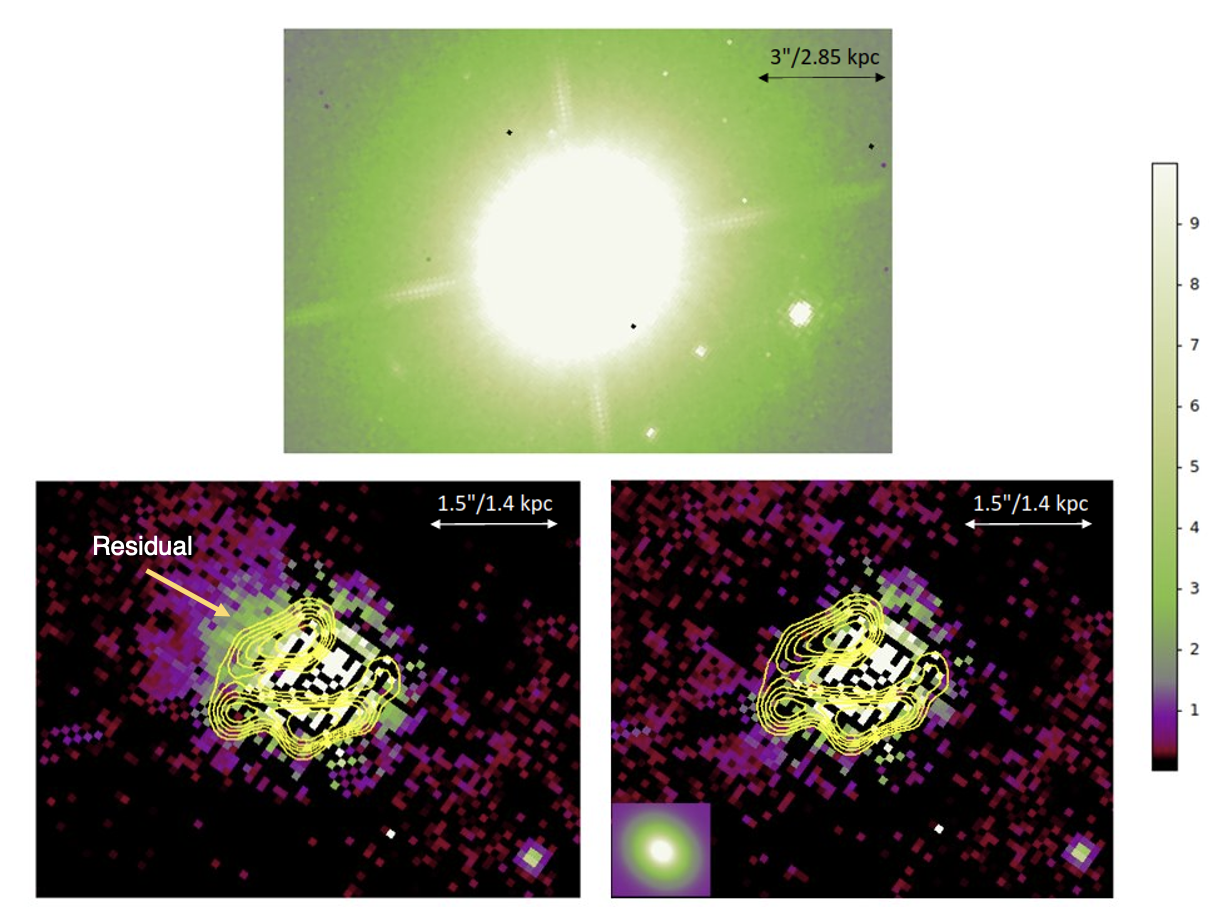}
  \caption{Top panel: Zoomed-in HST/WFC3 image of AP Librae before GALFIT model subtraction. Bottom panel: After subtraction of the S\'{e}rsic+PSF (left) and S\'{e}rsic+Expdisk+PSF (right) models, overlaid with ALMA Band 7 contours, where the sub-mm residual emission coincide with that in the IR. The used Expdisk model is shown in the small inlet inside the second figure. It is clear that the latter model allows better subtraction and is hence preferred.}
  \label{fig:hst1}
\end{figure*}

We used the publicly available GALFIT (GALaxy-FITting; \citealt{peng02}) routine to model the galaxy and central point source.  GALFIT takes in the source image and PSF as FITS file inputs and fits a user-chosen model of one or more components. All fitting was done using non-distortion-corrected \texttt{\_flc} images and the non-distortion-corrected PSF, where we used only a 300$\times$300 pixel subset of the image centered on AP Librae.  We used a simple model with a single S\'{e}rsic component, an exponential disk (to model the residual dust emission observed by ALMA) plus PSF to model the galaxy and point source in each of the four HST exposures. Addition of an exponential disk reduced the chi-square and subtracted residual emission which was otherwise visible with a simple Sersic+PSF model (see Figure \ref{fig:hst1} where we illustrate this difference). We utilized a custom mask in the fitting to avoid biasing the fit from either the jet or extraneous background objects and other artifacts. We masked out the jet region (using radio maps as a guide) as well as any bright source in the 300$\times$300 region.  The average (over each raw image) best galfit parameters are given in Table~\ref{table:galfit}.

\begin{figure*}[!t]
  \includegraphics[width=\linewidth]{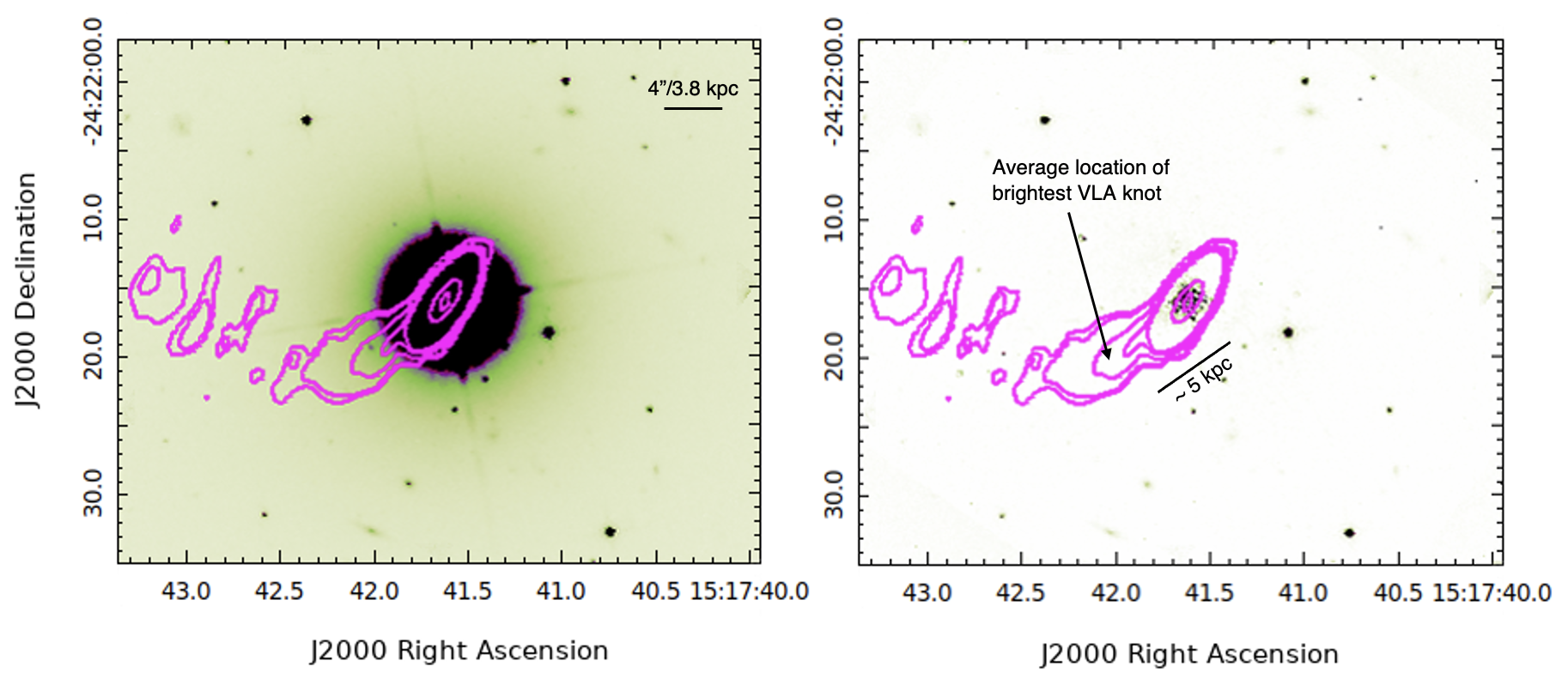}
  \caption{Drizzled HST/WFC3 images of AP Librae before (left) and after (right) GALFIT model subtraction, overlaid with VLA L-band contours, as in Figure~\ref{fig:aplib_radio}. The brightest VLA L-band knot location is marked by an arrow.}%The speck of emission outside the inner knot in both images is a background galaxy, whose contribution was neglected while computing the IR upper limit.}
  \label{fig:hst2}
\end{figure*}

After running GALFIT on each exposure to produce a galaxy/PSF subtracted distorted image, we then used AstroDrizzle and Tweakreg with standard settings to align the subtracted frames to produce a final subtracted and distortion-corrected image. To convert from count rate to flux units, we used the python {\tt PYSYNPHOT} package. We adopted a flat spectral slope through the F160W band and an extinction magnitude $E(B-V)$ of 0.1177 \citep{schlafly11}. A comparison of the resulting image stack using non-subtracted frames and subtracted frames is shown in Figure~\ref{fig:hst2}, with L-band VLA radio contours overlaid.

There is no indication of any jet-related flux by eye in the HST imaging after subtraction. We confirmed this and derived an upper limit for the IR flux density. Due to imperfections in the galaxy subtraction, we carefully sampled areas of the subtracted image corresponding to the same radial distance from the core (corresponding to $\sim$ 5 kpc, shown in Figure \ref{fig:hst2}) to derive our $5\sigma$ upper limit of 0.70 $\mu$Jy.

Further, using the value of the magnitude photometric zeropoint for HST/WFC3 as 28.2 (see HST/WFC3 handbook) in Equation 5 in \cite{peng02}, we obtained the total integrated flux density of the exponential disk model, or the IR dust emission, as $1.27$ mJy. The semi-major axis of this disk is $\simeq6$ pixels, and with a plate scale of $\sim0.12"/$pixel for WFC3, it turns out to be $\simeq0.7$ kpc, consistent with ALMA observations. It is imperative to note that the flux of this model is $\lesssim10^{-3}F_{sersic+psf}$, due to which the corresponding lower bound to this flux is poorly constrained. Observing the worsening of the fit, we hence choose a $50$ and $20\%$ lower and upper bounds respectively to the flux, implying $F_{ir}\simeq1.27^{+0.25}_{-0.65}$ mJy. However, we note that this has little to no impact on our conclusions.  %^{+0.10}_{-0.30}$ mJy.

\subsection{Fermi}

AP Librae has been detected by the \emph{Fermi}/LAT and is listed in the 4FGL-DR2 point source catalog as 4FGL~J1517.7-2422 with a detection significance of 95. We have analyzed over 12 years of continuous \emph{Fermi}/LAT observations to derive an average spectrum for this source, and have also derived the minimum flux state using the progressive-binning analysis method first presented in \cite{meyer14}. We briefly summarize the procedure here. \textit{Fermi}/LAT event and spacecraft data were extracted using a 7$^{\circ}$ region of interest (ROI), an energy cut of 50~MeV-300~GeV, a zenith angle cut of $90^{\circ}$, and the recommended event class and type for point source analysis (128 and 3 respectively).  The time cuts included all available \textit{Fermi} data at the time of analysis, with corresponding mission elapse time (MET) ranges of 239557417 to 626971271 (corresponding from 4 August 2008 to November 13 2020).

\begin{figure}[h!]
\centering
\includegraphics[width=3.2in]{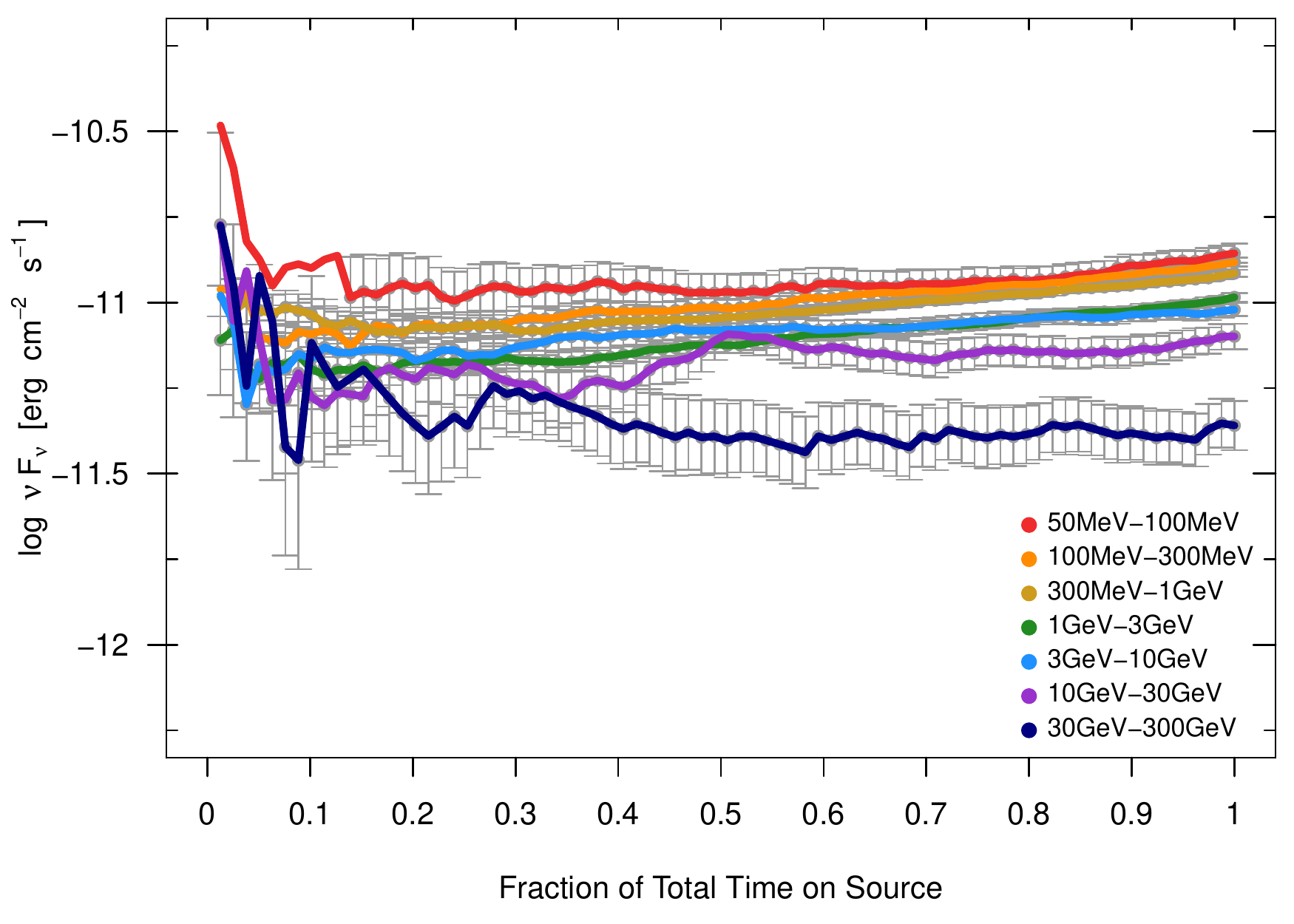}
\label{fig:fermi}
\caption{\emph{Fermi}/LAT flux versus total time on source (number of bins combined) for the progressive-binning analysis AP Librae, where the fluxes are for the seven energy bands used by the 4FGL-DR2 and listed in Table~\ref{table:fermi}.  }
\end{figure}

Using a model file populated with known 4FGL sources, we first generated a light curve by dividing the data into three week on-source time bins\footnote{The time bins were defined in terms of good time interval (GTI) time, corresponding to roughly 8 weeks in real time.} and running the pre-likelihood analysis tools on each time bin separately. In the maximum likelihood analysis for each time bin we model AP Librae as a power-law with a fixed photon index of 2.1, all other sources in the ROI similarly have the spectral shape (but not normalizations) fixed during the fitting.
We then used the combined-bin analysis procedure in order to determine the apparent minimum flux in each of the seven standard Fermi energy bands as used in the 4FGL catalog. The minimum and (12-year) average fluxes are given in Table~\ref{table:fermi}, along with the energy bin boundaries, central frequency, and the corresponding number of bins used to derive the minimum flux. The plot of flux versus number of bins combined for the seven energy bands is shown in Figure~\ref{fig:fermi}.

\begin{deluxetable*}{llcccc}[!ht]
\tablecaption{\label{table:fermi} Fermi Observations of AP Librae}
\tablecolumns{6}
\tablewidth{0pt}
\tablehead{
$\mathrm{E}_\mathrm{min}$ & $\mathrm{E}_\mathrm{max}$ & Log Freq. & N & $\mathrm{F}_\mathrm{min}$ & $\mathrm{F}_\mathrm{12yr}$ \\
 & & (Hz) &  & $\times10^{-12}$ erg\,s$^{-1}$\,cm$^{-2}$ & $\times10^{-12}$ erg\,s$^{-1}$\,cm$^{-2}$ }
\startdata
50 MeV	& 100 MeV	& 22.23	& 19	&  1.01$\pm$ 0.20	&  1.39$\pm$ 0.09 	\\
100 MeV	& 300 MeV	& 22.62	& 11	&  0.74$\pm$ 0.12	&  1.31$\pm$ 0.04 	\\
300 MeV	& 1 GeV  	& 23.12	& 15	&  0.82$\pm$ 0.06	&  1.22$\pm$ 0.03 	\\
1 GeV  	& 3 GeV  	& 23.62	& 4	&  0.60$\pm$ 0.13	&  1.04$\pm$ 0.03 	\\
3 GeV  	& 10 GeV	& 24.12	& 3	&  0.50$\pm$ 0.19	&  0.95$\pm$ 0.04 	\\
10 GeV	& 30 GeV	& 24.62	& 9	&  0.50$\pm$ 0.21	&  0.80$\pm$ 0.07 	\\
30 GeV	& 300 GeV	& 25.36	& 7	&  0.35$\pm$ 0.25	&  0.44$\pm$ 0.07 	\\
\enddata
\end{deluxetable*}
%: 50~100~MeV, 100~MeV-300~MeV, 300~MeV-1~GeV, 1~GeV-3~GeV, 3~GeV-10~GeV, and 10~GeV-30~GeV, 30~300~GeV. In all bands, the minimum flux level was consistent with the detection of a steady source. Further, the lightcurve of AP Librae is consistent with a steady source punctuated by outbursts, as further discussed in the next section.

\section{Results and Discussion}
\label{sec:3}
\subsection{Evaluating the IC/CMB Model}
\label{sec:iccmb}

\begin{figure*}[ht!]
\centering
\includegraphics[width=6in]{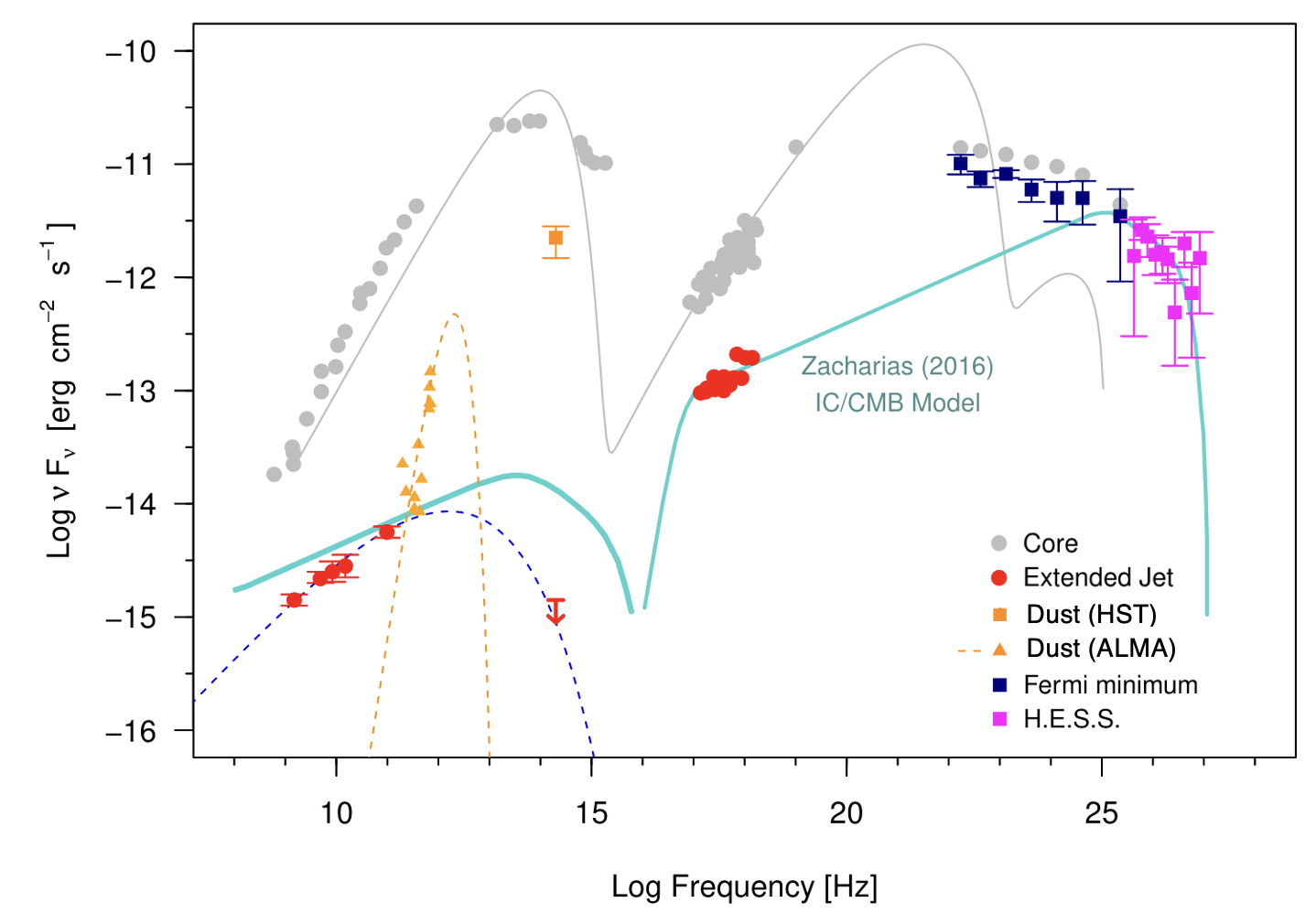}
\caption{The spectral energy distribution of AP Librae. The bright blazar core dominates over the kpc-scale jet from radio to X-rays and the data shown in gray are historical total source (i.e., core) fluxes. The corresponding gray solid line is a one-zone SED model for the core (see Section \ref{sec:icdust}). From our high-resolution VLA, ALMA, and HST imaging campaign we have been able to more accurately sample the synchrotron spectrum of the X-ray emitting jet (red data points and upper limit with phenomenological power-law+exponential cutoff model as a blue dashed curve). The infrared upper limit (in red) in particular rules out the IC/CMB model of \cite{zac16}, shown as cyan curves. In addition, a fit (orange-dashed) to the orange ALMA data points (uncertainties omitted to prevent clutter) has been indicated, which is a blackbody at temperature 25 K, having typical size $\sim$ 100 pc. The \textcolor{black}{estimated dust flux in the IR} is given as an orange box. }
\label{fig:sed_zach}
\end{figure*}

We present an updated spectral energy distribution (SED) for AP Librae in Figure~\ref{fig:sed_zach}, where all observations of the kpc-scale jet below X-ray energies as well as the \emph{Fermi}/LAT observations are newly presented in this paper. The historical total source fluxes (dominated by the blazar core) shown in gray are taken from \cite{zac16} (hereafter Z16) and/or NED. We have used red data points for the kpc-scale jet fluxes and the IR upper limit, and orange triangles for the jet+dust fluxes measured by ALMA for the frequencies above 100 GHz (which are clearly dominated by the thermal component). An orange square point has been used for the estimated flux of the (presumed) dust disk residual from our IR imaging. The X-ray spectrum for the kpc-scale jet is also shown as red points \citep[from][]{kauf13}. In the GeV-TeV band we show the \emph{Fermi}/LAT 12 yr average flux in dark gray and the minimum LAT fluxes in dark blue. Finally the H.E.S.S. TeV flux points are shown in magenta \cite[from][]{hess15}. In this section, we will primarily focus on the CMB seed photon field and the validity of the IC/CMB model for the TeV emission. In following sections, we discuss the dust seed photon field in detail and its contribution to the broadband SED.

While we are able to separately resolve the core and the kpc-scale jet at lower frequencies, in the GeV-TeV regime the observations shown in Figure \ref{fig:sed_zach} are for the entire source and it is not clear which zone of emission dominates. The light gray and cyan model curves are taken from Z16. The thick cyan curve is their synchrotron model for the large-scale jet and the thin cyan curve the corresponding IC/CMB emission proposed to explain the kpc-scale X-ray and (total) TeV emission. The required synchrotron spectrum under this model is now clearly very discrepant considering the new data points. The radio spectrum is significantly harder than assumed in Z16, while our IR upper limit for the jet is far below the model curve. In Z16 there is also only one radio flux point for the kpc-scale jet -- an L-band (1.4 GHz) flux \textcolor{black}{density} of 0.20 Jy reported by \cite{cas99}, which is significantly higher than our value. This is due to \cite{cas99} reporting the total jet flux from a composite map including a D-configuration observation, which recovers flux from the outer (non-X-ray emitting) jet on scales $\lesssim200''$. Instead we have confined our measurements to the $4''$ scale jet from which X-ray emission is detected.

\begin{figure*}[ht!]
\centering
\includegraphics[width=6in]{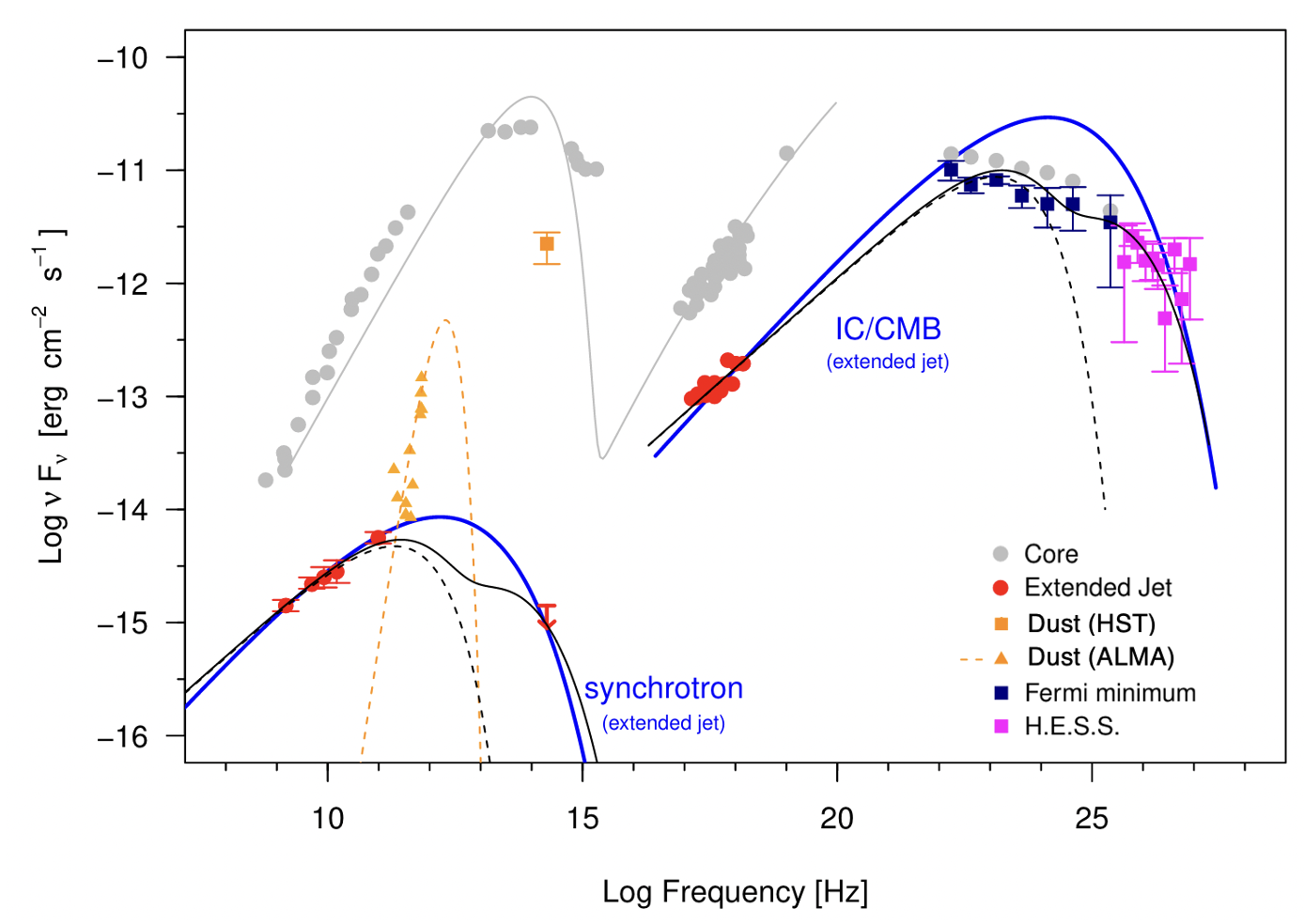}
\caption{
\label{fig:aplib_sed2} Data points are the same as in Figure~\ref{fig:sed_zach}. Three different models invoking the IC/CMB mechanism for the kpc-scale X-ray emission are shown. The ``naive'' model is shown in blue, which matches well the radio-IR points but greatly overpredicts the GeV band. The ``Fermi-obeying" model is shown as dashed black curves. This single-component model slightly underpredicts the ALMA 100 GHz flux of the jet and cannot explain the TeV emission. The solid curve shows an ad-hoc two-component synchrotron model which can explain not only the X-ray emission on kpc scales but also the VHE flux. %The gray solid line has been truncated to avoid confusion with other models for the VHE emission.}
}
\end{figure*}

While it is clear that the specific IC/CMB model of Z16 is now ruled out, it is worthwhile to consider whether an IC/CMB model for the kpc-scale X-ray emission could still be compatible with the GeV-TeV observations, if not fully explain them. In Figure~\ref{fig:aplib_sed2} we show the same data as in the previous figure, and consider different possible IC/CMB scenarios given the constraints on the kpc-scale synchrotron emission. In the ``naive'' scenario, we fit a phenomenological powerlaw with exponential cutoff model to the radio through IR data, shown as a thick blue line. The key parameters that govern the synchrotron and correspondingly the IC-spectral shape are the Doppler factor $\delta$ ($=[\Gamma(1-\beta\cos\theta)]^{-1}$ where $\beta$ is the jet bulk speed, $\Gamma$ is the bulk Lorentz factor and $\theta$ is the angle between the jet and our line of sight) and the magnetic field $B$ \citep[e.g.,][]{georg06}, where the \textcolor{black}{IC spectral shape} can derived from simple shifting of the synchrotron spectrum in frequency and luminosity.  A B/$\delta$ value of $\sim10^{-7.5}$ allows the corresponding IC/CMB component to match the X-ray flux level with a spectral index ($F_\nu\propto\nu^{-\alpha}\sim\nu^{-0.72}$) in good agreement with the \cite{kauf13} observations ($\alpha=\Gamma_{ph}-1\simeq0.8$ where $\Gamma_{ph}
\simeq1.8$ is the photon index in $N(\nu)\propto\nu^{-\Gamma_{ph}}$). However, such a model greatly overpredicts the \emph{Fermi}/LAT spectrum. Indeed, to obey the \emph{Fermi}/LAT minimum flux and still explain the X-rays, the synchrotron spectrum must peak (in $\nu F_\nu$) at 100-200 GHz, as shown by the dashed curves. For this scenario, we had to soften the radio spectral index to a degree that leads to an under-prediction of the ALMA band 3 flux (the highest frequency red data point shown) by about 20\%. Considering the difficulty of properly accounting for the jet flux given the thermal contamination and extremely dominant core in the ALMA bands, we cannot rule this scenario out on that basis alone. However, this model clearly cannot explain the VHE emission.

The only IC/CMB model which can satisfy the current observational constraints below $10^{15}$ Hz and explain the TeV emission without violating the \emph{Fermi}/LAT minimum is a somewhat contrived two-component model, shown as a solid black line in Figure~\ref{fig:aplib_sed2}. Here we have simply added a second (phenomenological) synchrotron spectrum, peaking just below $10^{14}$\,Hz to the previous dashed-curve spectrum. Such a model does appear to match well the GeV-TeV spectrum of AP Librae, however we have no direct evidence for the existence of the second synchrotron component, which would require a separate electron energy distribution with slightly higher maximum energy compared to that producing the radio.

\subsection{Properties of the dust spectrum}
\label{sec:dust}
Based on the core-subtracted imaging shown in Figure \ref{fig:alma}, the residual ALMA dust emission appears to arise from a disk possibly being viewed at a small inclination (or ``face-on"). This is the first detection of large-scale dust emission in a BL Lac and is in marked contrast with studies that have otherwise failed to detect any dust emission from a pc-scale torus in a large sample of BL Lacs \citep{plot12}.  Detailed modelling of the dust structure and emission \citep[e.g.,][]{fritz06, nenk02, stalev12} is beyond the scope of this paper. We employ a very simple cylindrical model (with thickness and height) to model its structure and thereby the emission spectrum as in Section \ref{sec:alma}, \textcolor{black}{where we determined that the approximate radius $R$ of the disk is $0.65$ kpc, and the inferred scale height using $H\simeq R/5$ is $0.13$ kpc.} The inner radius $r_{in}$ of the disk must connect with the pc-scale torus (if present) and for the purpose of modeling we neglect this as $r^2_{in}\ll R^2$. Assuming simplistically the disk thermally emits at a single temperature, the total luminosity of the disk would be given by $L=(2\pi R(R+2H))F=(14\pi R^2/5)F$, where $F$ is the total flux integrated over all frequencies emitted by the disk. If we use the blackbody flux $F=\sigma T^4$ and luminosity distance $D_L=215$ Mpc, we find that the ALMA fluxes match a blackbody of temperature $T=25$ K and $R=0.10$ kpc. The fitted value of $R$ is shorter than an effective observed size ($\sqrt{RH}\simeq0.40$ kpc) as most of the flux comes from the central part of the dusty disk. While there are a range of blackbodies of different temperature and spatial size which are consistent with our data, we chose the above parameters since they are consistent with our imaging and SED modelling (Section \ref{sec:icdust}) and slight differences do not affect our results. \textcolor{black}{However, as a very conservative estimate, if $r_{in}$ is actually larger than assumed/observed, e.g., $r_{in}\sim0.5R$, $L$ is only reduced by $\sim$15\%. The fact that this is insensitive to the outcome of the paper is further exemplified in Appendices \ref{sec:app1} and \ref{sec:app2}. }
%Figures \ref{fig:sed_zach} and \ref{fig:final_sed} show the final SED of the dust spectrum in orange.

From HST imaging, we only have one observation in the IR, which equals a flux value $F_{ir}\simeq 1.27$ mJy at $10^{14.30}$ Hz. The size of the dust disk, as inferred indirectly from model-fitting using GALFIT, was $R\simeq 0.5$ kpc, consistent with that obtained from ALMA. The 25 K blackbody that describes the ALMA observations grossly underestimates this IR flux (Figure \ref{fig:final_sed}), implying we require a better model to describe the total dust spectrum from sub-mm to the IR. Hence we represent the HST flux with a Heaviside spectral flux density centred at $10^{14.30}$ Hz with width equal to that of the F160W filter. We use both the above models of the dust seed photon fields in the upcoming subsections to explain the origin of TeV emission using inverse Compton scattering.

\subsection{Revisiting inverse Compton models}

The discussion in the Section \ref{sec:iccmb} disproved a simple one-zone IC/CMB model for the origin of the VHE emission but did allow a double-synchrotron model to explain the broadband SED of AP Librae. Our objective in this section is to make indicative fits to the spectral energy distribution (SED) of AP Librae discussing the possibilities of different seed photon fields as a source of TeV emission through external Compton by hot electrons in the jet/core.

\subsubsection{Seed photon fields within the kpc-scale X-ray emitting jet}

We consider the case of the kpc-scale X-ray emitting jet of AP Librae \citep{kauf13}, which moves with bulk Lorentz factor $\Gamma$ \textcolor{black}{(and speed $\beta$)} and most emission occurs within $\sim 10$ kpc from the sub-kpc core, with the bright knot(s) lying at $z_0\lesssim5$ kpc (\cite{kauf13} and Figure \ref{fig:aplib_radio}). If we assume the average angle of the kpc-scale jet to our line-of-sight ($\theta$) to lie between 5 and 15 degrees, following spectral modelling by \cite{hervet2015} and \cite{zac16}, it implies that the maximum de-projected jet length is 114 kpc, while the brightest knot lies at $z_x=\langle z_0/\sin{\theta}\rangle\simeq$ 35 kpc for $z_0=4$ kpc. In the following paragraphs, we discuss the importance of different seed photon fields for inverse Compton emission from the kpc-scale jet.
%Let us call the latter distance $r_{x}=35$ kpc, signifying the distance at which maximum energy dissipation occurs along the kpc-scale jet.

\paragraph{Radiation from the blazar core}

Let the blazar core be defined by a size $r_c=1$ pc. Radiation from this core is relativistically beamed into an opening angle $\lesssim1/\Gamma_{core}$, considering the sub-pc jet to be moving with a bulk Lorentz factor $\Gamma_{core}$ (and a Doppler factor $\delta_c$). For an approximate value of the deprojected distance along the kpc-scale jet (with $\theta\simeq10^{\degree}$) $z_{deproj}=1\unit{kpc}/\sin10^{\degree}=5.75\,\unit{kpc}\gg r_c$, it is safe to assume that radiation from the core illuminates the kpc-scale jet directly from behind, implying the nuclear radiation is Doppler de-boosted in the kpc-scale jet frame. Following \cite{stawarz03}, the observed energy density of the blazar radiation in the frame of the energetic electrons in the kpc-scale jet can then be written as follows (primed coordinates refer to the kpc-scale jet frame):

\begin{equation}
U'_{core}=\frac{L_{core}(\theta=0)}{4\pi z^2c}\frac{1}{4\Gamma_{rel}^2}\Bigg\rvert_{z=z_{deproj}}
\end{equation}

where $L_{core}$ is the isotropic co-moving blazar luminosity, $z$ is the de-projected distance along the jet from the jet apex and $\theta$ is the angle between the line of sight of an observer in the kpc-scale jet frame and the blazar radiation. \textcolor{black}{$\Gamma_{rel}=\Gamma_{core}\Gamma(1-\beta_{core}\beta)$ which is the relative bulk Lorentz factor between the jet and the core. $1/\Gamma^2_{rel}$ converges to 1 when $\Gamma_{core}\sim \Gamma$, while deviates to $\ll1$ when the core and jet speeds are very different. Since it is very unlikely they will have similar speeds, we can safely assume we will have deboosting of the core emission. While \cite{stawarz03} consider the blazar core to be stationary in the frame of the kpc-scale jet ($\Gamma_{rel}=\Gamma$), we have relaxed the assumption here.} Using conservative estimates of $\Gamma_{core}\simeq\delta_{core}$ and $L_{core}\simeq10^{43}$ ergs$/$s ($\simeq L_\gamma$ from the SED) for the blazar core we can evaluate $L_{core}(0)$ using \cite{stawarz03}. $L_{core}(0)=L_{core}(2\Gamma_{c}/\delta_c)^3$, or $L_{core}(0)=10^{44}$ ergs$/$s. Therefore, the energy density of the blazar radiation in the frame of the kpc-scale jet is $U'_{core}=8.0\times10^{-14}/\Gamma_{rel}^2$ ergs cm$^{-3}$.

\paragraph{Cosmic Microwave Background}

The CMB is isotropic and it is preferentially boosted in the direction of the fast-moving kpc-scale jet. At a redshift of $z=0.048$, the CMB energy density is given by \citep[see e.g.,][]{tav00,georg06}

\begin{equation}
U'_{cmb}=aT_{cmb}^4(1+z)^4\Gamma^2\simeq4.8\times10^{-13}\Gamma^2\unit{ergs\,cm^{-3}}
\end{equation}
where $a=7.53\times10^{-15}$ cgs units is the radiation constant, $T_{cmb}=2.73$ K is the temperature of the CMB at $z=0$. Therefore it directly follows that \textcolor{black}{$U'_{core}/U'_{cmb}\simeq0.1/\Gamma_{rel}^2\Gamma^2$}, or the energy density of the nuclear radiation $U'_{core}\ll U'_{cmb}$ and can be neglected in the frame of the kpc-scale jet. We also do not consider \textit{synchrotron radiation from the kpc-scale jet} as a dominant source of seed photons since the corresponding energy density is orders of magnitude lower than the CMB.

\paragraph{Large-scale dust emission (ALMA)}

%Generally the torus in AGN extends only up to a few parsecs in size (CITE) and is a rich reservoir of photons for the extended jet. However, since the extent

The radiation from a possible (currently undetermined) pc-scale dusty torus, if present in AP Librae, will have a minimal energy density as it reaches the kpc-scale jet since it will be Doppler de-boosted and hence will lose relevance. Therefore, in this section we will only consider the ``outskirts" of the molecular torus, or the 500 pc-scale photon field imaged using the ALMA and HST.

The minimum temperature blackbody that the ALMA data points describe has a temperature of 25 K and is $\sim 100$ pc in size. Generally, for the case of the pc-scale torus, the dust photons are preferentially de-boosted in the frame of the kpc-scale jet since they are very far away and hence only illuminate the jet electrons from behind. In this case, however, \textcolor{black}{if one uses} the observed size $R\sim0.65$ kpc from ALMA imaging, one needs to derive the exact energy density as a function of distance along the jet. Simplistically assuming the dusty disk has spherical symmetry for calculating the isotropic luminosity and that the jet is dominantly illuminated by the surface of the disk facing it, we derived the energy density of the kpc-scale dusty disk in the co-moving jet frame in Appendix \ref{sec:app2}. It shows the dependence of the energy density with increase in distance along the jet, from the jet apex. It naturally gives rise to beaming at very low distances compared to the size of the disk and de-beaming at distances $z\gg R$ along the jet.

%an approximate de-projected jet location of the brightest knot $z_x\sim35$ kpc and $\Gamma\simeq 5$ for the kpc-scale jet, we find $R/z_x\simeq0.5/\Gamma^2$, which is a sizeable fraction of $1/\Gamma^2$. Therefore we assume that the dust radiation is Doppler boosted in the frame of the kpc-scale jet. Further if we approximate the dust disk to be spherically symmetric at a large distance from the jet apex along the de-projected jet (e.g., $R\ll z_x$),

%the energy density of this dust photons in the frame of the kpc-scale jet would be given by:

\paragraph{Large-scale dust emission (HST)}

This emission is centred at the observing frequency, i.e., $\nu_0=10^{14.30}$ Hz and would constitute a source of a seed photon field for the kpc-scale jet. The width of the F160W filter for WFC3 is $\Delta\lambda\sim0.4\mu m$, which transforms to $\Delta\nu=10^{13.72}$ Hz. The flux density $F_{ir}=1.27$ mJy, which implies the total integrated flux would simply be given by:

\begin{equation}
F=\int^{\infty}_0 F_{ir}\Theta(\nu-\nu_0+\Delta\nu)\Theta(\nu_0+\Delta\nu-\nu)d\nu=F_{ir}(2\Delta\nu)
\end{equation}

where $\Theta$ is the Heaviside step function assuming equal transmission through the entire filter. At most, this assumption underestimates the total flux as we are using a monochromatic spectrum. Therefore the luminosity of the dusty disk at the source would simply be given by $L_{ir}=F(4\pi D_L^2)$ where $D_L=215$ Mpc is the luminosity distance to the source. Using the given values, we obtain $L_{ir}\simeq 6.0\times10^{42}$ ergs$/$s. The expression for the energy density due to the infrared-emitting region is the same as in Appendix \ref{sec:app2}. In the upcoming sections, we will discuss this further.

For all our calculations in the Appendix, we use a spherically symmetric assumption for the dusty disk to calculate the isotropic luminosity. Assumption of spherical symmetry of the dusty disk at the de-projected kpc-scale jet is at variance with works that follow better approximations; see for example \cite{stalev12}, \cite{fritz06}, \cite{nenk02} for a discussion on pc-scale tori emission models and \cite{drouart12} for a thorough observational study of pc-scale torus-jet alignment in radio galaxies. In this case, our ALMA imaging confirms that our dusty disk is very likely being viewed at a low inclination. While the jet maybe aligned at a different angle which implies that a part of the disk preferentially illuminates the nearer side of the jet, this mis-alignment between the jet and the disk/torus needs meticulous treatment and is out of the scope of this paper.

%In this work, although we only use a very rough assumption of spherical symmetry for the isotropic luminosity, we rigorously estimate the energy density in the co-moving jet frame as described in Appendix \ref{sec:app2}.

\subsubsection{Dust photon fields in the frame of the pc-scale core}

The parsec and sub-pc core/jet receives most of the radiation from the inner part of the dusty disk and can be assumed to be approximately isotropically illuminated for this work. We will find that this assumption will not change our final conclusions.

%However, a proper treatment of dealing with energy densities of the BLR and the torus in the co-moving jet frame have been given in \cite{hay12}.

%We currently cannot observationally constrain the inner extent of the disk and hence this will be used as a free parameter for modelling the spectral energy distribution.

\subsection{Evaluating the IC-dust model}
\label{sec:icdust}
%The discussion in the previous section disproved a simple one-zone IC-CMB model for the origin of the TeV emission and invoked a combination of two synchrotron components to explain the broadband SED of AP Librae. In this section, we explore the possibility of electrons in the pc-scale core and the kpc-scale jet IC-scattering dust photons to produce the $\gtrsim100$ GeV emission.
%However, due to the complexity of our problem, we employ a rigorous modelling approach.

We here use a simple one-zone homogeneous leptonic model to understand the spectral energy distribution (SED) of AP Librae. Our code models both the synchrotron and inverse Compton (IC) emission, where we have also implemented Klein-Nishina effects at higher energies. We consider a region of size $R$ containing a homogeneous magnetic field $B$ moving relativistically with a bulk Lorentz factor $\Gamma$. Radiation from the region is beamed with Doppler factor $\delta$. Electrons with relativistic energies following a power law $Q(\gamma)=N_0\gamma^{-p}$ are injected into the region at timescales much shorter than the light crossing time. The electron energy distribution $n(\gamma,t)$ is self-consistently evaluated at later times using the Fokker-Planck equation:

\begin{equation}
\frac{\partial n(\gamma,t)}{\partial t}+\frac{n(\gamma,t)}{t_{esc}}=\frac{\partial}{\partial \gamma}[\dot{\gamma}n(\gamma,t)]+Q(\gamma)
\end{equation}

where $t_{esc}$ is the average time an electron spends in the emission region and the first and second terms on the right hand side represent cooling and injection respectively. For our purpose, we use the \textit{total} energy density of the CMB and the dust (sub-mm and IR) for simultaneous inverse Compton scattering by electrons in the same emission region. Although we use energy densities strictly following the results in Appendix \ref{sec:app2}, we boost the frequency (by $\Gamma$) for simplicity. We will shortly find that this choice is immaterial.
\textcolor{black}{The set of parameters of the broad dust spectrum required to produce the VHE emission has been tabulated in Table \ref{tab:par}, with $U_{cmb}=4.2(1+z)\times10^{-13}\Gamma^2 \unit{ergs\,cm^{-3}}$ and $h\nu_{cmb}/m_ec^2=(1+z)\times10^{-9}$.}

\begin{figure*}[ht!]
\centering
\includegraphics[width=6in]{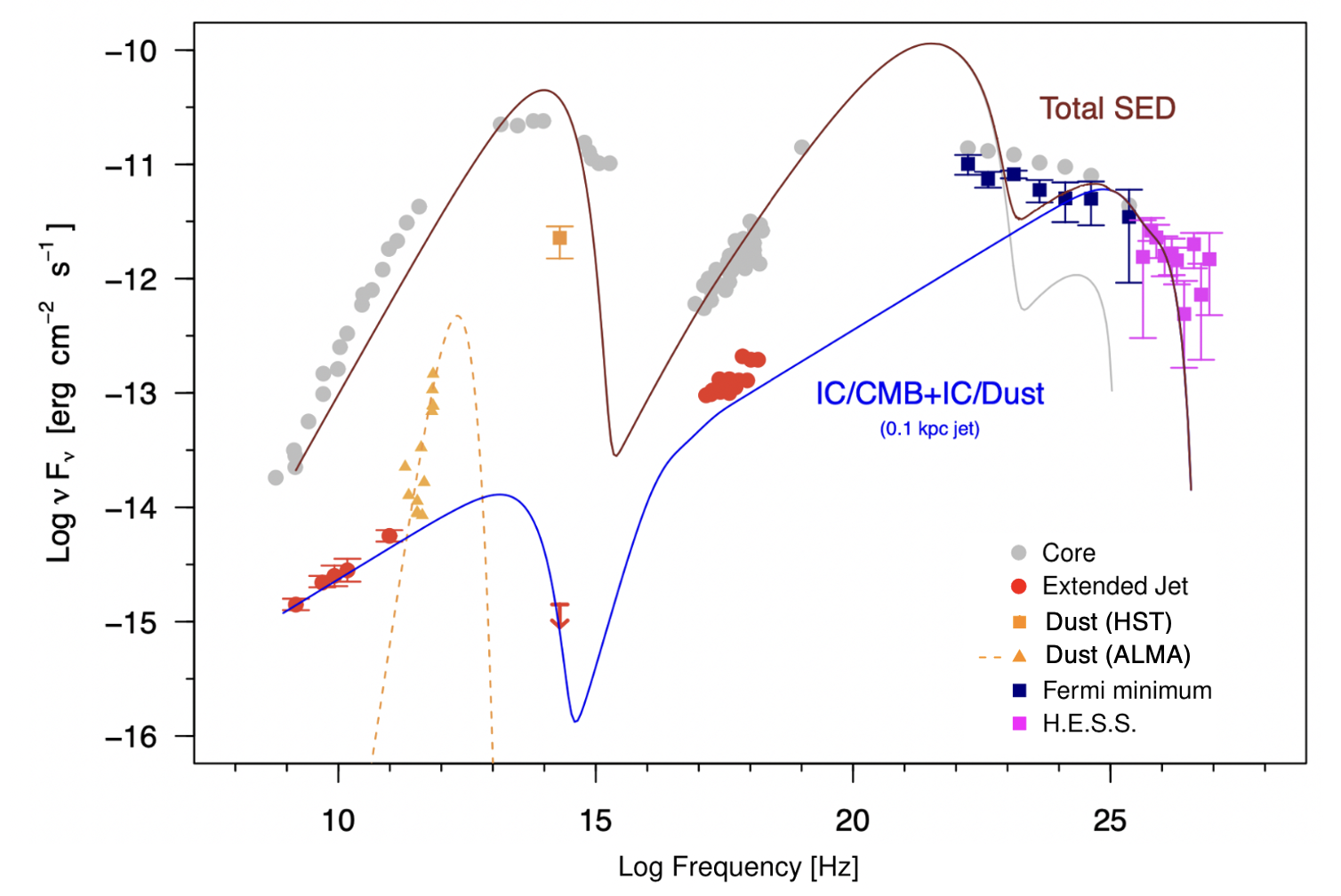}
\caption{Data points are as in Figure \ref{fig:sed_zach}. Orange fitted line is the blackbody fit to the ALMA fluxes for T=25 K and size 100 pc. The radio-TeV SED of the extended jet has been shown in blue solid line. The gray line is the total SED of the core, including multi-order synchrotron self-Compton and IC/dust. It is clear that the total SED (maroon solid line) describes the GeV and most part of the TeV emission, where the latter arises from IC/CMB and IC/Dust in the extended jet.}
\label{fig:final_sed}
\end{figure*}

%Observationally, the cirumnuclear disk/torus extends through few hundred parsecs and is a reservoir of seed photons for the sub-pc core as well as the extended kpc-scale jet. %Although underconstrained, we found that found two blackbodies of temperature 15 K and 500 K to well describe the broad IR spectrum, as shown in Figure \ref{fig:icdust}.

%Following discussion in the previous section, the most relevant photon fields for electrons in the extended $>0.5$ kpc jet are the CMB and the sub-mm to IR dust photon fields.

The final SED has been shown in Figure \ref{fig:final_sed}.  The orange data points describe the ALMA and the HST dust fluxes and the orange dashed line represent the 25 K blackbody fit with 100 pc size, as discussed earlier. The gray line is the spectrum of the blazar core and the total (core+jet) SED is given in brown. We find that the total inverse Compton spectrum due to IC/CMB, IC/sub-mm (inverse Compton scattering of sub-mm dust photons) and IC/IR (inverse Compton scattering of IR dust photons), shown in a solid blue line, produces the entire GeV-TeV spectrum. The IC/CMB majorly produces the low-energy end of the VHE spectrum (like Figure \ref{fig:aplib_sed2}) while the sub-mm dust photons are IC-scattered to produce the TeV emission. Since we used a very conservative estimate of the luminosity for the infrared photons, the IC spectrum is poorly sensitive to the IC/IR photons. Hence in the coming sections we will mainly discuss the sub-mm photon field. However, since we expect the dust emission to be bright through a range in IR frequencies, a complete observation in the IR may allow us to constrain the VHE better. In addition, the total inverse Compton spectrum clearly underproduces the X-rays by $\sim30\%$. It is not clear if the X-rays are due to IC/CMB+IC/Dust or they are produced due to synchrotron emission in a different emitting zone in the extended jet. In fact, the origin of X-rays in extragalactic jets is a long-debated question \citep[e.g.,][]{harris06,georg06,meyer15,breiding17}, and also remains open in AP~Librae. \textcolor{black}{The GeV spectrum is only approximately described by our model. The detection of small flares in the Fermi band over the last decade may indicate some contamination from the core. However, since we are only using an indicative fit to constrain the TeV emission, rather than a simultaneous fitting of all data, such discrepancies are considered only minor. It is possible that an additional pc-scale jet model similar to \cite{hervet2015} may be better able to fit the observations, although at the cost of significant parameter degeneracies.}

\begin{figure}[ht!]
\centering
\includegraphics[width=\linewidth]{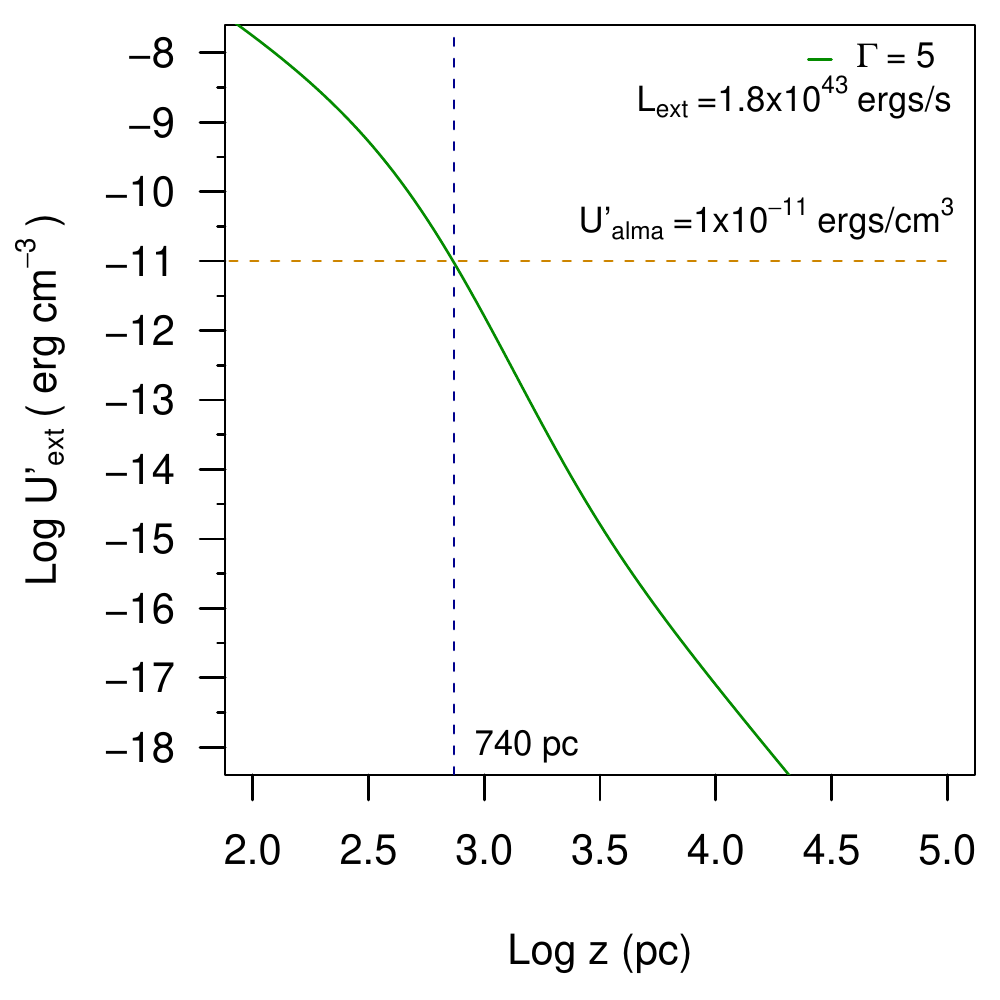}
\caption{Figure shows the external photon (dust) energy density (in a solid green line) as a function of de-projected distance $z$ along the jet. Using the energy density from the SED model ($U'_{alma}$ in a dashed yellow line), we calculate the $z$ at which the ALMA seed photons are inverse Compton scattered to produce the TeV emission.}
\label{fig:submm}
\end{figure}

Now that the SED model has the energy densities of the seed photon fields \textit{required} to produce the VHE emission, it is imperative to determine the corresponding distance along the jet where the required sub-mm energy density would hold, allowing us to determine the location of VHE emission using results from Appendix \ref{sec:app2} and check if it is physically feasible. From Figure \ref{fig:submm} we find that the dust photons (and the CMB) are inverse Compton scattered to GeV-TeV majorly at $r_{diss}\sim$ 740 pc along the jet, which translates to $\sim0.07$ kpc in projected scale, at the HST/ALMA resolution limit. In addition, we adopt $\Gamma=5$ for the jet and we have beamed the frequency of the seed photons. Incorporation of the beaming pattern of the dust from the Appendix would at most reduce it by $(1-\beta r_{diss}/\sqrt{R^2+r_{diss}^2})\simeq0.4$, or $\nu_{seed}$ would effectively be boosted by $\Gamma=2$ instead of 5, which would reduce the maximum IC-scattered frequency. This discrepancy can be resolved by noting the fact that the peak of the sub-mm blackbody is essentially unconstrained, therefore allowing us to increase the temperature and thereby the frequency of the seed photon field, although that may reduce the fitted size of the disk. Moreover, it is also not clear if there is only one blackbody since we also observe the residual emission in the mid-IR, which implies that the true peak frequency is underestimated. This might result in underestimation of the emitting size (compared to the observed) when a single blackbody is used for the fit, as we saw in Section \ref{sec:dust}. Without filling the gap in the sub-mm to mid-IR observations, none of the above can be constrained and hence, for the broader implications of this work, the assumption of boosting the seed photon frequency is plausible.

\begin{figure*}[!t]
\centering
\includegraphics[width=7in]{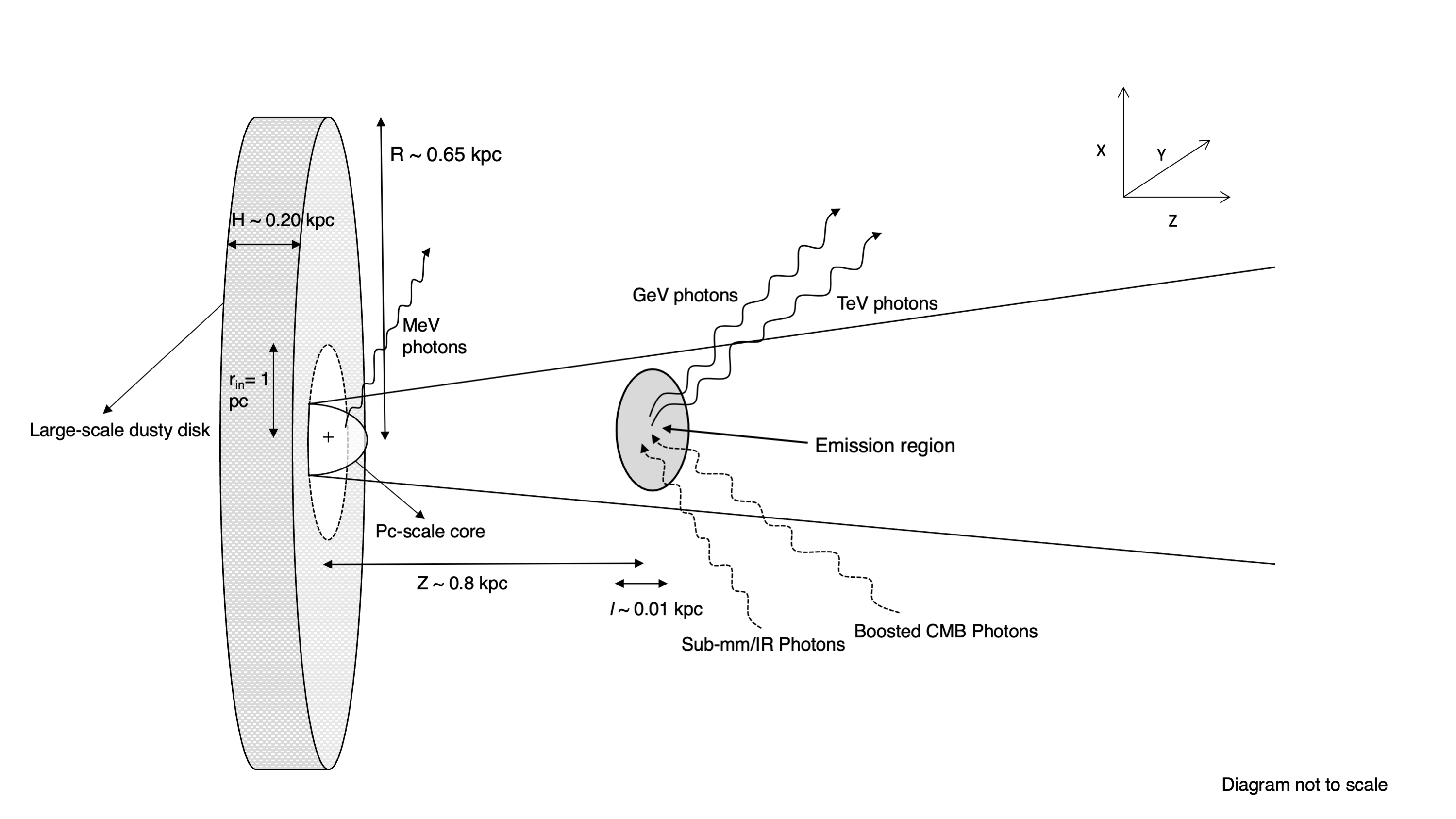}
\caption{An illustration of the assumed morphology, emission mechanisms and locations as determined from observations and SED modelling. Figure shows the basic model of the dusty disk surrounding a pc-scale core (with the jet apex marked with a cross), with the disk having approximate minimum radius as $r_{in}$. The kpc-scale jet is shown to the right with the emission region size $l\simeq0.01$ kpc obtained from the SED model (Table \ref{tab:par}). The GeV and TeV emission are produced at $\sim$0.8 kpc (deprojected) along the sub-kpc-scale jet by IC/CMB and IC/sub-mm scattering respectively. }
\label{fig:sketch}
\end{figure*}

In Figure \ref{fig:sketch} we show an illustration of the structure of the source, the different emission mechanisms and their locations. It shows a basic cylindrical model of the dusty disk covering a pc-scale core, aligned perpendicular to the extended jet. The GeV-TeV emitting region is mostly in the projected $<0.1$ kpc (or de-projected $\lesssim0.8$ kpc) portion of the jet, as expected from the beaming pattern of the dust seed photons in the frame of the jet. The radio-emitting region is spread through $\sim0.01$ kpc, corresponding to the size of the emission region, but the fluxes in the radio SED are dominated by different parts of the extended jet. The figure is broadly consistent with the results of our work.

%This can be resolved firstly, by considering the fact that the emission region is $\sim1$ kpc in size and hence encompasses both the radio emitting region as well as the TeV. Secondly, as shown in Figure \ref{fig:ku}, the VLA 15 GHz image of the source shows a very bright inner jet $<0.2$ kpc which dominates the 15 GHz flux used in the SED.

%The dominant radio emitting locations, from VLA observations, are considerably dissimilar, spread through $<0.2$ kpc to $>5$ kpc, which is consistent with the total size of the emission region in our model.

%In any case, the sub-kpc jet is slow ($\Gamma=5$) and the kpc-scale jet will be even slower, so for spectral models with emission region located in the latter this discrepancy will decrease.

\begin{deluxetable}{lll}[h!]
\tablecaption{Physical parameters of the SED model for the core and the extended jet emission. \textcolor{black}{$R_{em}$} refers to the size of the emission region and $L_e$ is the injected electron power. The energy densities $U$ have additionally been defined in Equation \ref{a3}. The parameter $z_{jet}$ refers to the de-projected distance from the jet apex where external Compton emission is dominant. It has been determined using the corresponding energy density in Figure \ref{fig:submm}. The corresponding parameters for the core have considerable freedom and the dissipation locations are typically $\lesssim10$ pc, occurring in the pc-scale jet, which we have shown in Figure \ref{fig:sketch} but not in an energy density plot.}
\tablecolumns{5}
\tablewidth{0pt}
\label{tab:par}
\tablehead{Parameter & Core & Extended Jet}
\startdata
$\delta$ (Doppler factor) & 15 & 8.5 \\
$\Gamma$ (Bulk Lorentz factor) & 10 & 5 \\
$\theta$ & $5.2^{\degree}$ & $4.7^{\degree}$ \\
\textcolor{black}{$R_{em}$} (pc) & 0.003 & 13 \\
$t_{esc}$ (l/c) & 0.45 & 8.0  \\
$p$ & 1.40 & 2.45 \\
$\gamma_{min}$ & 10 & 20  \\
$\gamma_{max}$ & $6.5\times10^3$ & $8\times10^5$ \\
B (G) & 0.08 & $2.3\times10^{-6}$ \\
$L_e$ (ergs$/$s) & $10^{43}$ & $3.65\times10^{44}$ \\
$U_{cmb}$ (ergs cm$^{-3}$) & $4.41\times10^{-13}\Gamma^2$ & $4.41\times10^{-13}\Gamma^2$ \\
$\nu_{cmb}$ (m$_ec^2/$h) & $1.05\times10^{-9}\Gamma$ & $1.05\times10^{-9}\Gamma$  \\
$U_{alma}$ (ergs cm$^{-3}$) & $10^{-7}$ & 1$\times10^{-11}$  \\
$\nu_{alma}$ (m$_ec^2/$h) & $1.2\times10^{-8}\Gamma$ & $1.2\times10^{-8}\Gamma$ \\
$z^{alma}_{jet}$ (pc) & $24.3$ & $740$\\
$U_{hst}$ (ergs cm$^{-3}$) & $2\times10^{-7}$ & 5$\times10^{-11}$ \\
$\nu_{hst}$ (m$_ec^2/$h) & $1.6\times10^{-6}\Gamma$ & $1.6\times10^{-6}\Gamma$ \\
$z^{hst}_{jet}$ (pc) & -- & -- \\
\enddata
\end{deluxetable}

%From Figure \ref{fig:ir} we find that the TeV emission is dissipated at 380 pc along the jet, which translates to $\sim0.05$ kpc in projected scale, just below the resolution of ALMA/HST/VLA.

The location of the VHE emitting region in the jet is not unexpected. At a large distance along the jet away from the dusty disk, the seed photons will be Doppler de-boosted, thereby making IC/dust irrelevant. Hence the above estimates of locations $r_{diss}<1$ kpc are intuitively expected. However, it is imperative to reconcile these findings with the location of synchrotron-emitting region. The radio and the X-rays mainly arise from the kpc-scale jet. The same set of the electrons that produce the radio synchrotron inverse Compton scatter dust and CMB photons to produce the GeV-TeV emission. Although the projected location along the jet where this dissipation occurs is $r_{diss}\sin\theta<0.1$ kpc and is at variance with the observed radio emission location observed using VLA 1.4 GHz, which is in the projected kpc-scale jet, the dominating radio emission region changes observed location with change in frequency. Hence, a sub-kpc location of emission from radio to TeV using a one-zone emission model is plausible for the broader implications of this work. We do not claim a precise determination of the location of the emission region since it is model-dependent, but we can say the dissipation region must not be $>1$ kpc away in projected distance from the central engine, or the seed photon energy density would be too low to produce the VHE emission. Accurate determination of the emission location requires other diagnostic tests \citep[e.g.,][]{harvey20} and we assume for this work it lies within the 0.1 kpc projected-jet and is at the resolution limit for the HST/ALMA. In addition, the size of the emission region from the modeling is $R_{em}\sim10$ parsecs. This is consistent with a sphere which fits inside a jet which has a cross-sectional dimension $ r_{diss}/\Gamma\simeq150\unit{pc}>R_{em}$. Using $\delta=8.5$ (Table \ref{tab:par}), a model-dependent predicted variability time scale would be $t_{var}\simeq3.5$ years. Ideally, since the CMB is static, one would not expect a variability signature. However, since we have multiple seed photon fields, variability may be complex. Future monitoring of the extended jet from this source will allow a strong constraint on the true size of the dominant emission region, if it is variable.

Emission from the blazar core, in contrast, was unperturbed by the presence of the seed photon fields. This is expected because the magnetic energy density is orders of magnitude larger than the maximum seed photon energy densities ($U_B\simeq5\times10^{-4}\unit{ergs/cm^3}\gg U_{ext}\sim10^{-7}$ ergs$/$cm$^3$, extrapolating $U_{ext}$ in Figure \ref{fig:uext} to $\sim100$ pc, or 10 pc in projected scale) and $\gamma^2_{max}\nu_{ir}\simeq10^{22}\unit{Hz}\ll\nu_{GeV}\gtrsim10^{23}$ Hz.

With a large-scale dusty structure as observed, it is also imperative to consider $\gamma-\gamma$ pair production on the dust photon fields as possible disruption of the $>$ GeV radiation within the inner jet. Since even the ``highly beamed" jet may encounter this large disk since it is on the $\sim$ kpc scale, one cannot rule out the $\gamma-\gamma$ opacity using relativistic beaming arguments. The threshold energy of the seed photon required for pair production by a $E_{ext}=1$ TeV photon is $E_{thr}=2(m_ec^2)^2[E_{ext}(1-\mu)]^{-1}\gtrsim0.25\unit{eV}$, or $\nu_{thr}=10^{13.8}$ Hz for head-on collision ($\mu=-1$), which is $\sim$ IR. However, the energy density of the observed IR photons is too low to be considered relevant in this case. Even if we conservatively assume the IR photon field is a blackbody spread through the size of the ALMA imaged structure (0.5 kpc), the corresponding ``compactness" parameter is $\bar{l}\simeq U\sigma_T\bar{R}/m_ec^2$ ($\bar{R}$ is the size of the absorption region and U is the IR energy density; e.g., \citealt{rous11,bottcher16}), or $\bar{l}\simeq 2.5\times10^{-6}\bar{R}_{pc}\ll1$ using a rough estimate of U($z<50$ pc) $\simeq10^{-6}$ ergs$/$cm$^3$ from Appendix \ref{sec:app2}. This implies that the large-scale dust is essentially optically thin to the TeV radiation from the pc-scale core to the kpc-scale jet.

Finally, we note that the $\sim10$ TeV emission is underproduced by our model. It is possible the high-energy TeV emission is unphysical and an artifact of inaccurate EBL corrections. However, since we do not have enough multi-wavelength observations, we cannot rule out a high IR energy density spread through mid to near-IR, inverse Compton scattering of which may produce the high TeV. Note that Klein-Nishina effects will begin to appear at these seed photon energies.

\subsection{Jet energetics and composition}

%We have discussed this further in the final section.
The question of jet composition, and the total power thus carried, is still not fully settled. A purely leptonic jet contains light electron-positron pairs, as produced purely in a black hole spin-powered jet built in the black hole magnetosphere \citep{bland77}, while a purely hadronic jet is produced from the accretion disk and contains electron-proton pairs \citep{bland82}. In FSRQs, due to the presence of very strong photon fields in the vicinity of the pc-scale jet, the jet must have a minimum proton content to prevent significant deceleration due to Compton drag on the $e^--e^+$ pairs and minimize heavy annihilation signatures which are otherwise not observed. For HBLs, since external photon fields are either not observed due to beaming or are intrinsically weak, the proton content is difficult to constrain \citep{madej16}. However, AP Librae is a low-synchrotron peaking BL Lac (LBL) that has a sub-kpc scale dusty disk, suggesting the possibility that it is a so-called `fake' BL Lac  \citep[e.g.][]{keen21} with strong accretion and therefore pc-scale seed photon fields. Photons from the large-scale dust can be important for both the pc-scale and the kpc-scale jet. However, none of these can be used to constrain the proton content using present observations so we will dedicate this subsection to making rough estimates of the minimum power demanded by our model and its general plausibility.

The Eddington luminosity for AP Librae is $L_{edd}\simeq3\times10^{46}\unit{ergs/s}$. The injected electron power (or power in ``hot" electrons), from Table \ref{tab:par}, is simply $P_e=3.65\times10^{44}$ erg/s. A crude estimate of the radiative power $P_r$ (which is the power spent by the electrons to produce the observed luminosity), due to the anisotropic nature of inverse Compton scattering, can be obtained from \cite{ghis10} as $P_r\simeq3\Gamma^4\delta^{-6}L_\gamma\simeq10^{40}\unit{ergs/s}$ where $L_\gamma\simeq10^{42}\unit{ergs/s}$ is a proxy for the total luminosity. $P_r$ is therefore an observational limit. Additionally, the magnetic field power $P_B\simeq\pi R_{em}^2\Gamma^2U_Bc\simeq10^{39}\unit{ergs/s}\ll P_e$, implying the jet is most likely matter dominated. The most important questions, however, are regarding the fraction of the ``cold" electrons, number of electron-positron pairs and number of protons per pair. We will employ the simplest approach of power estimates by measuring that for a purely leptonic (pairs only) and for a purely hadronic (e-p) jet.
For a leptonic jet, if we assume an acceleration mechanism only energizes fraction $\eta_e$ of the electrons, the total power in the leptons would then be $P_{e^-e^+}=P_{cold}+P_{hot}=\frac{1-\eta_e}{\gamma_{min}\eta_e}P_e+P_e\simeq5.2\times10^{44}\unit{ergs/s}$ for $\eta_e=10\%$ and  $\langle\gamma\rangle\simeq\gamma_{min}$, where the latter provides an upper bound to the power as ideally $\langle\gamma\rangle\simeq[(p-1)/(p-2)]\gamma_{min}>\gamma_{min}$ for $p>2$. Therefore the total jet power in the purely leptonic case would be $P_l=P_r+P_{e^-e^+}+P_B\simeq5.5\times10^{44}\unit{ergs/s}\simeq0.02L_{edd}$. For a hadronic jet, if we assume fraction $\eta_e$ of the electrons are heated by tapping the inertia of the protons and fraction $\eta_p$ is the acceleration efficiency, $P_e\simeq \eta_p P_p$. Therefore, the total hadronic power $P_{ep}=P_{cold}+P_{hot}=\frac{1-\eta_e}{\gamma_{min}\eta_e}P_e+P_e/\eta_p+P_e\simeq4.1\times10^{45}\unit{ergs/s}$ for $\eta_p\sim10\%$, or the total jet power $P_h\simeq4.2\times10^{45}\unit{ergs/s}\simeq0.14L_{edd}$. However, in addition to $\gamma_{min}$, the above estimates are sensitive to $\eta_e$ and $\eta_p$, which are unconstrained in general. For a more conservative estimate of $\eta_e\simeq1\%$, $P_l\simeq2\times10^{45}\unit{ergs/s}\simeq0.07L_{edd}$ and $P_h\simeq5.6\times10^{45}\unit{ergs/s}\simeq0.2L_{edd}$. The extended jet power must interpolate between $P_l$ and $P_h$ and is sub-Eddington at all times in contrast to previous literature on AP Librae \citep{hervet2015,petr17}. We also note that the power estimates for the core are likely to be similar to or lower than that for the extended jet since the hot electron power $L_{e,core}\lesssim10L_{e,jet}$, $\gamma_{min}$ is half that of the jet, the magnetic power  $P_B\simeq10^{41}\unit{ergs/s}$ (Table \ref{tab:par}) and $P_{r,core}\simeq\Gamma^2\delta^{-4}L_{\gamma}\lesssim10^{41}\unit{ergs/s}\gtrsim10P_{r,jet}$ (the SSC emission is isotropic in the co-moving frame). Therefore the total (core+jet) power may at most be double that of the jet and hence will still be sub-Eddington. However, it is not clear to what extent $\eta_e$ and $\eta_p$ depend on the type of acceleration and if it will stay the same for a purely leptonic and a purely hadronic jet. For the case of AP Librae, further constraints on the cold particle fraction can be obtained through future studies of faint diffuse emission in the jet, where particles are not energetic enough and bulk Compton signatures may prevail (e.g., see \citealt{georg05,mehta09} for the case of PKS 0637-752).

In the light of the above discussion, we also note that the maximum apparent speed observed for the pc-scale jet is $\beta_{app}\simeq6$ \citep{lister19}, which implies $\theta_{max}\simeq19^{\degree}$, consistent with Table \ref{tab:par}. Assuming the ``pattern" speed represents the actual bulk flow, $\Gamma_{min}\simeq6$ for the pc-scale core, consistent with $\Gamma_c=10>\Gamma_{min}$ obtained from SED modelling. For the larger-scale jet, proper motion studies do not yet exist. However, as ALMA observations suggest, a large gas reservoir must exist within the few kpc of the galactic nucleus and may cause sufficient mass loading (see e.g., \citealt{perucho14} and next section) and/or Compton drag \citep[e.g.,][]{sikora96} if the jet is light. Hence a deceleration from $\Gamma_c=10$ to $\Gamma_{jet}=5$ appears reasonable.

\subsection{Possible origin of the dust emission}

The maximum extent of a torus such that it is bright in the mid-IR due to AGN-heating and is under the influence of the black hole will be given by $r<r_{max}=GM_{BH}/\sigma_b^2$ where $M_{BH}$ is the mass of the black hole and $\sigma_b$ is the velocity dispersion of stars in the spherical bulge of the galaxy \citep{alex12}. Using $M_{BH}=2.5\times10^{8}$ M$_\odot$ \citep{woo05} in the classic $M-\sigma$ relation from \cite{geb00}, we obtain $\sigma_{b}\simeq240\unit{km/s}$. This gives us $r_{max}\simeq20$ pc, which is more than 10 times smaller than the structure detected in our ALMA imaging. This larger-scale dust is cold
%and is bright mainly in the far-IR. Both of these observations imply the observed torus is
and clearly outside the influence of the AGN, but may plausibly indicate hot dust on parsec scales (i.e., a classical molecular torus).
%and is mainly driven by galactic dynamics or possible starburst heating.
%In the following paragraphs, we will discuss the possible origin of the circumnuclear dust and the extent to which it fits into the radio-loud AGN unification paradigm.

\begin{figure}[!t]
  \includegraphics[width=\linewidth]{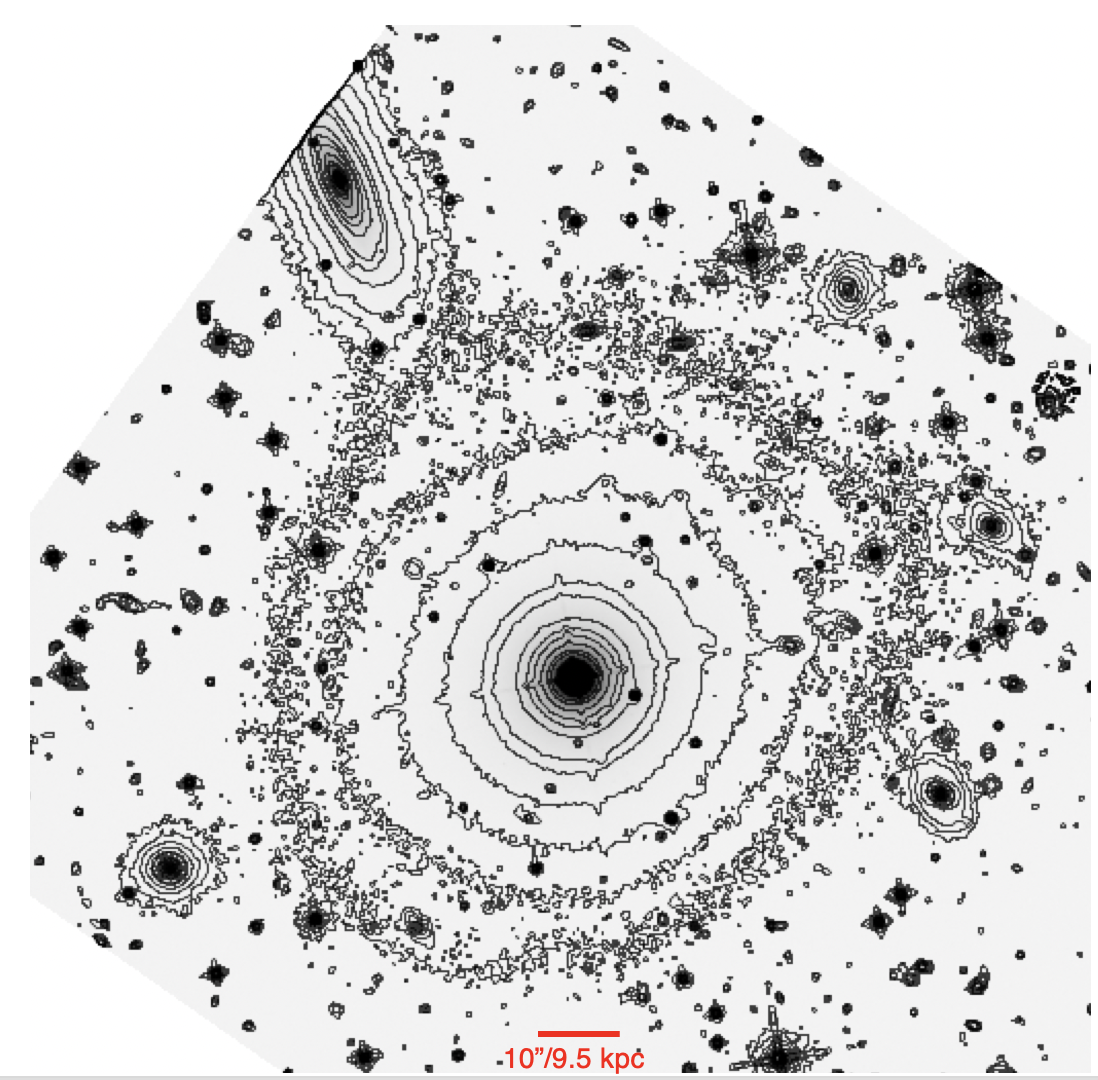}
  \caption{HST-WFC3 image of AP Librae showing it in an interacting pair of galaxies, confirmation of what was first observed by \cite{stickel94} and \cite{pesce94}. The large object is AP Librae and its host galaxy while approximately $\sim65''$ to the NE, lies a lenticular galaxy 2MASX J15174385-2421212. The outermost contours of AP Librae show distinct elongation in the direction of the secondary galaxy, suggestive of a tidal interaction.}
  \label{fig:int}
\end{figure}

The elliptical host of AP Librae has a close  companion galaxy, 2MASX J15174385-2421212, located $65''$ ($62$ kpc) to the NE and at the same redshift \citep{pesce94}.
Figure \ref{fig:int} shows the closely interacting pair in our 1.6 $\mu$m image, with visible tidal distortions in the outer NE contours/isophotes of the AP Librae host.
It is a long-standing claim that AGN activity is a result of major mergers (e.g., \citealt{hopkinsquat10}, but see also \citealt{lambr21}), and this appears to be an even stronger claim for radio-loud AGN \citep{chiaberge11}. The environment of radio-loud AGN has been observed to more likely have a close companion instead of normal gas-rich galaxies \citep[e.g.,][]{ellison11,tadh16}, with FRIIs more likely than FRIs to be associated with an interacting pair.   The AP Librae host is considerably brighter (by 1.5-2 orders of magnitude) than the companion and is presumably far more massive. This is consistent with the findings of \cite{ellison11}, where they find in case of unequal-mass interacting pairs, the higher mass primary galaxy is more likely than the lower mass secondary to host an AGN.

AGN activity is thought to be triggered by large quantities of dust and gas from scales $>10$ kpc being driven into the nucleus, to scales $\sim$ pc \citep[e.g.,][]{hopkinsquat10}.
Assuming a patch of dust/gas $\sim0.5$ kpc away from the AGN is much lighter than the black hole ($\sim10^8$ M$_\odot$), an upper bound on the time taken to fall into the black hole, or the free-fall time, is found to be $t_{ff}\lesssim 10^7$ years. This is roughly at par with a typical AGN lifetime. For dust/gas masses at most $\sim10^{10}$ M$_\odot$, consistent with galaxy-scale dust emission (see e.g., \citealt{hickox18}), $t_{ff}\simeq10^6$ years. %This implies that it is definitely
It is possible that the large-scale dust that we see is the result of the galaxy interaction driving gas into the pc-scales and feeding the black hole, resulting in a radio-loud AGN with a relativistic jet. %A rigorous confirmation of the above is not possible with present ALMA/HST observations.
Only a detailed highly-resolved spectroscopic/imaging ALMA-JWST (James Webb Space Telescope) study of the inner torus would reveal vital details regarding the dynamics of outflow/inflow from kpc to pc scales.

%In addition, it is imperative to mention that we do not know the composition of the excess circumnuclear emission that we are referring to as ``dust". It emits as a blackbody but it can be cold molecular gas possibly acting as a reservoir for major inflows in the smaller scales. Hence, only for the purpose of the merger/interaction driven black hole fuelling, we assume it to be either dust or gas.

The confirmation of a major interaction triggering AGN activity should  be visible in other observations of this system, like evidence of enhanced star formation and molecular inflow. Indeed, detailed photometric observations of AP Librae have revealed its host galaxy to be much bluer than average \citep{baxter87} in its nuclear regions ($<10$ kpc, observed during low AGN activity) and redder at larger radii, in apparent contrast with both typical ellipticals and early-type spirals. Thus it appears plausible that the sub-kpc dust (we see in the far and mid IR (Figure \ref{fig:alma})) is heated by the nuclear starburst launched by the galaxy interaction rather than the AGN.

%This is further confirmed by the observation of no far-IR residual emission above $\sim0.7$ kpc away from the nucleus, consistent with higher star formation in the nuclear ($<100$ pc) regions and extended (sub-kpc) far-IR emission from starlight-heated dust. Kpc-scale mid-IR emission is majorly contaminated by light from the host galaxy and hence we did not find any large-scale residue after galaxy subtraction using the HST (Figure \ref{fig:hst1}). The starburst heating is most prominent in the nuclear regions and hence we detected the excess mid-IR emission at $\lesssim0.5$ kpc.

It appears that this is the first time a torus or circumnuclear dusty disk has been observed in a BL Lac object, though dusty disks have been observed in several FR I radio galaxies \citep{sparks00}. %This is atypical of pc-scale tori generally known but it is naturally expected that this large-scale torus extends to the tens of parsec scales and possibly further since we have an active AGN.
\cite{plot12}, using $\sim$ 100 WISE-selected weakly-beamed BL Lacs, failed to detect the presence of a pc-scale torus in the entire sample. This may be due to the fact that low-power BL Lacs are poor accretors \citep[e.g.,][]{hardcastle20} and hence dust is poorly heated, in contrast to FSRQs\footnote{and presumably `fake BL Lacs' which are really FSRQs with their broad lines buried under the relativistic jet} where there has been observational confirmation of dust tori \citep{castig15}.
%However, AP Librae is an unusual BL Lac, with a very broad SED and showing clear features of tidal interaction with a secondary galaxy. In addition, we may have a torus from sub-kpc to pc-scales, where the signature of the latter may be overshadowed by beamed emission from the AP Librae jet.
If there is a pc-scale molecular torus in AP Librae, we also expect a strong accretion signature. An accretion disk `big blue bump' (BBB) is not clearly visible in the SED, though \cite{hervet2015} and \cite{zac16} assign the apparent flattening of the core spectra in the optical-UV to a BBB.
%Either this is insignificant, or a strong accretion signature is overwhelmed by relativistic beaming, in which case the latter picture is consistent with a torus at pc-scales.  In addition,
The emission line spectra of AP Librae \citep[e.g.,][]{disney74, peterson76, rodgers77} show variable emission line intensities with equivalent widths $\lesssim2$\AA. It is unclear if this reflects a change in the gas supply near the core or if it is simply due to inherent variability in the competing beamed non-thermal radiation.

%Therefore, the features observed in AP Librae cannot be readily generalized to other BL Lacs since it is most likely a very specific kind, with features largely dependent on fortuity rather than being representative of the entire source class. Interestingly, however, another BL Lac H1517+656, sharing similarity with AP Librae (but not detected with Fermi-LAT), seems to be the remnant of a merger that possibly triggered the radio source \citep{odowd05}.

\section{Conclusions}
\label{sec:conc}

%As previously noted by \cite{san15}, \cite{hervet2015} and \cite{zac16}, simple one-zone leptonic SSC emission from the core, both first-order and second-order, fail to reproduce any TeV and most of GeV emission from AP Librae. In the GeV regime, the predicted SSC emission under-predicts the observed Fermi fluxes and is also too soft (see, e.g., Figure 4 of \cite{san15}, or Figure~2 of \cite{zac16} and \cite{hervet2015}). \cite{hervet2015} consider a blob-in-a-jet scenario, where they predict the TeV emission to arise from inverse Compton scattering of the large-scale jet synchrotron seed photons by relativistic electrons in the blob near the base of the jet. While the model does reproduce the total SED, it does so at the expense of having many free parameters. \cite{petr17} ascribed synchrotron radiation due to secondary electron-positron pairs formed from photo-hadronic interactions as the mechanism for TeV emission. The site of emission is the core in this case and they simulate corresponding variability signatures in the TeV that would be detectable in the future by the Cherenkov Telescope Array (CTA).

AP Librae is a unusual BL Lac object with a high-energy spectral component ranging over 9 decades in energy. As previously noted by \cite{san15}, \cite{hervet2015} and \cite{zac16}, simple one-zone leptonic SSC models for the blazar core (both first-order and second-order) fail to reproduce any TeV and most of GeV emission from AP Librae.
%In the GeV regime, the predicted SSC emission under-predicts the observed Fermi fluxes and is also too soft (see, e.g., Figure 4 of \citealt{san15}, or Figure~2 of \citealt{zac16} and \citealt{hervet2015}). \cite{hervet2015} consider a blob-in-a-jet scenario, where they predict the TeV emission to arise from inverse Compton scattering of the large-scale jet synchrotron seed photons by relativistic electrons in the blob near the base of the jet. While the model does reproduce the total SED, it does so at the expense of having many free parameters. \cite{petr17} ascribed synchrotron radiation due to secondary electron-positron pairs formed from photo-hadronic interactions as the mechanism for TeV emission. The site of emission is the core in this case and they simulate corresponding variability signatures in the TeV that would be detectable in the future by the Cherenkov Telescope Array (CTA). However, all these studies predicted very different jet energetics and composition.
In this work, we have detected circumnuclear dust of $<$1 kpc extent around the core of AP Librae, which we find to be a plausible of source of seed photons that can be upscattered to higher energies by energetic electrons in the kpc-scale jet. In contrast, based on our improved sampling of the synchrotron-emitting jet spectrum, the VHE emission cannot be produced by a simple kpc-scale IC/CMB model as previously suggested \citep{zac16}. Our preferred explanation for the origin of VHE emission is a combination of inverse Compton scattering of the CMB and dust photons by a single population of electrons in the sub-kpc jet. The jet power required in this scenario is significantly sub-Eddington, in contrast to most alternative models.

At VHE, a kpc-scale origin could be ruled out if high amplitude fast variability is detected (e.g., by the Cherenkov Telescope Array). Because of the scale of the resolved jet and the steadiness of the CMB and possibly of the dust on timescales of hundreds of years, the VHE emission resulting from IC processes in the extended jet should be non-variable. Although a few small-scale flares have been observed with Fermi, it is quite possible that these are due to variability in the core which likely contributes to the flux at GeV energies.

Further clarity on the dynamics of the dust, and thereafter the source of VHE emission from AP Librae would be gained with new, deep IR and sub-mm observations to better constrain the synchrotron emission of the jet (current observations up to $\sim$ 100 GHz do not probe the peak). \textcolor{black}{It is also unclear if the ALMA observed dust and the IR-detected residual are intrinsically connected, which can only be verified with deep far and mid-IR observations.} Spectroscopic observations with ALMA and JWST will also provide insights into the inflow/outflow dynamics or molecular and ionized gas, which can help us to understand triggering of AGN activity. In principle, deep observations in the far-IR with JWST could strengthen or rule out the extended jet as being the source of the TeV emission.
%since the Fermi-band has contributions from the core and IC spectra of both the CMB and the dust, no conclusions can be drawn. TeV variability, if present, might be less complicated to interpret if it is a result of IC/CMB and IC/Dust. However, it is grossly unclear to what extent the dust emission might vary and only through combined multi-wavelength imaging and spectroscopic campaigns can such mysteries be unravelled.

\vspace{2cm}

\section{Acknowledgment}

This paper is based on observations made with the NASA/ESA Hubble Space Telescope obtained from the Space Telescope Science Institute, which is operated by the Association of Universities for Research in Astronomy, Inc., under NASA contract NAS 5–26555. This work was supported by the HST-GO grant 15175.

This paper makes use of the following ALMA data: ADS/JAO.ALMA\#2017.1.01411.T, 2017.1.00568, 2017.1.00995, 2017.1.00258, 2017.1.01583, 2017.1.00963, 2017.A.00047, 2017.1.00239, 2013.1.00244, 2017.1.01555, 2017.A.00047, 2017.1.00337, 2017.1.00023, 2017.1.01555. ALMA is a partnership of ESO (representing its member states), NSF (USA) and NINS (Japan), together with NRC (Canada), MOST and ASIAA (Taiwan), and KASI (Republic of Korea), in cooperation with the Republic of Chile. The Joint ALMA Observatory is operated by ESO, AUI/NRAO and NAOJ. The National Radio Astronomy Observatory is a facility of the National Science Foundation operated under cooperative agreement by Associated Universities, Inc.

\textit{Facilities}: VLA, ALMA, HST, Chandra, Fermi, H.E.S.S.

\textit{Software}: CASA \citep{petry12}, SAO DS9 \citep{ds9}, GIMP \citep{gimp}.

\appendix

\section{ALMA Imaging of the Dust Emission}
\label{sec:app1}

\textcolor{black}{
The suggestion that the dust emission is continuous is not directly clear from or proven in Figure \ref{fig:alma}.
%As mentioned, a pseudo-symmetric effect can be due to over-subtraction of the core emission. This is generally not taken into account since in most radio-loud AGN, no strong isotropic extended emission around the core exists. It is mainly in the form of bright jet components distributed in an anisotropic manner at some distance from the core, like Figure \ref{fig:aplib_radio}. However, in this case we detect extended emission \textit{around} the core in the sub-mm frequency bands and it
It is a priori unclear if the best source structure is a point source atop a uniform disk or if the disk has a large inner radius. In this section, we demonstrate the validity of the assumptions used in the paper.}

\begin{figure}
    \centering
    \includegraphics[width=0.65\linewidth]{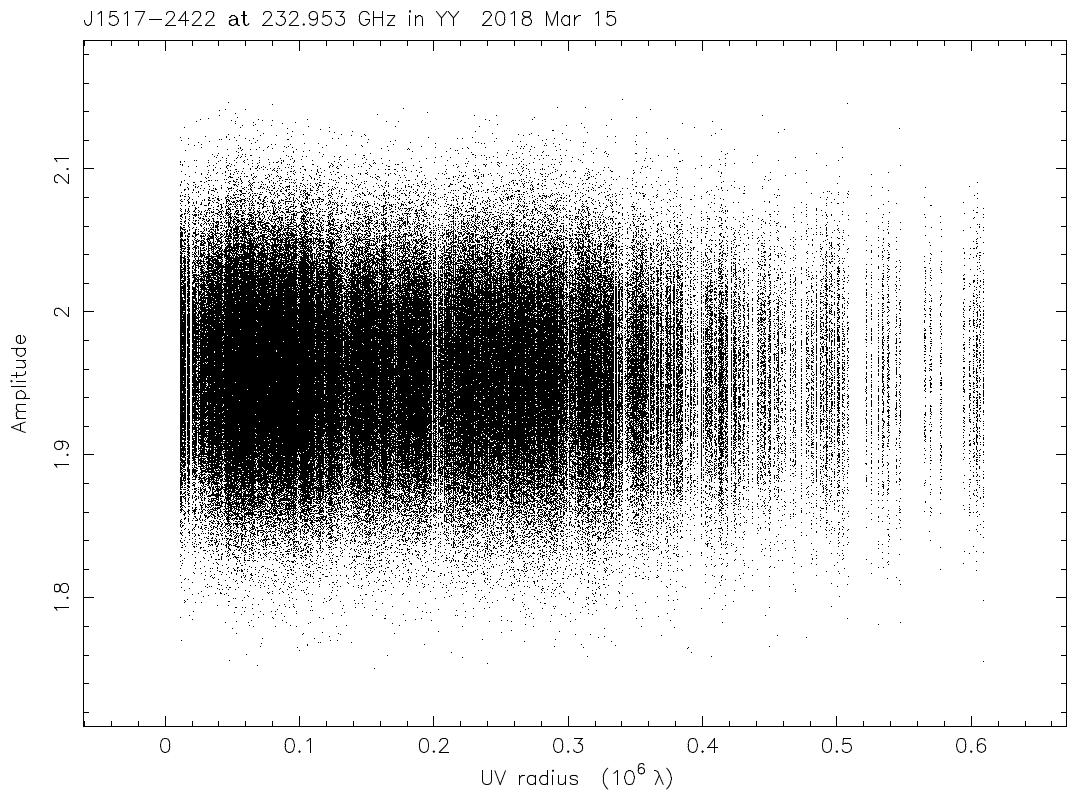}
    \caption{Visibility Amplitude v/s UV radius for the ALMA 232 GHz observation. The $u-v$ plane is visibly well-sampled. The amplitude is almost constant, demonstrating that the source is very core-dominated. The scatter is due to antenna gain errors.}
    \label{fig:uvcov}
\end{figure}

\begin{figure*}
    \centering
    \includegraphics[width=\linewidth]{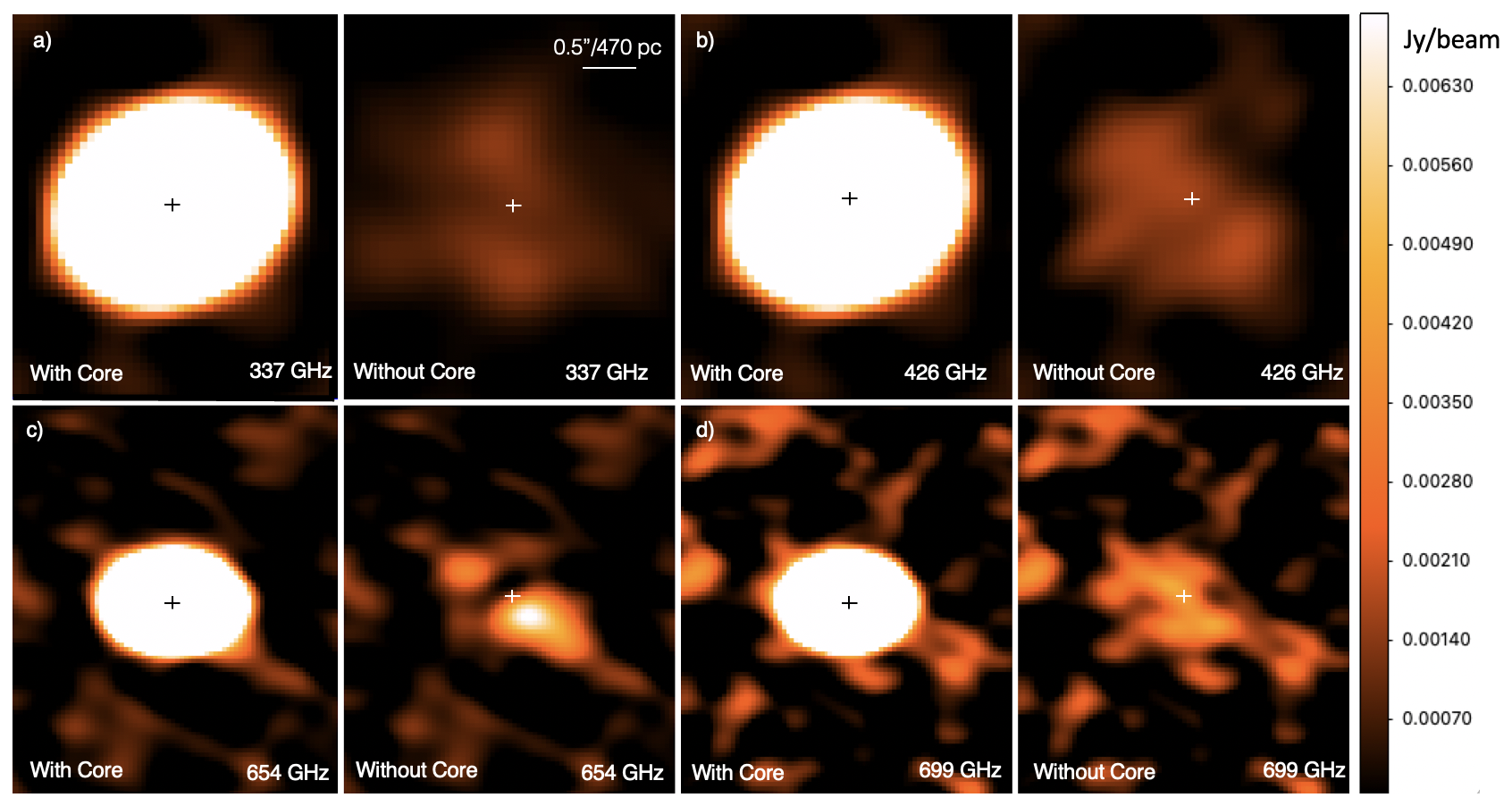}
    \caption{All the panels from (a) to (d) show the core ALMA images and the core-subtracted \textit{model residuals} for the frequency bands in Figure \ref{fig:alma}. The beam size being large, the extended emission (jet+dust) is barely visible in (a) and (b). The core-subtracted images show \textit{almost} continuous emission in all the observations. The core/peak flux is marked with a white/black cross.}
    \label{fig:alma2}
\end{figure*}

Generally in radio interferometry, in cases where the source is not observed long enough or multiple antennas have dropped out, large gaps in $u-v$ coverage can create false symmetric structures (or `side-lobes'), which must then be handled carefully (e.g., ignored during self-calibration). In all of the ALMA data used in the work, the total observing times were $\gtrsim20$ minutes and the $u-v$ coverage was dense (example visibility amplitude versus $u-v$ distance shown in Figure \ref{fig:uvcov}). Even if we assume the symmetric structures are due to the only gaps in the $u-v$ coverage (or even amplitude errors), their brightness/strength must follow the brightness of the core. However, from Table \ref{table:vla_alma} and Table \ref{tab:par}, the radio core spectrum is mainly flat ($\alpha\sim0.2$) while the brightness of the extended structure increases sharply with frequency following a quasi-blackbody spectrum. This further rules out any false positives regarding association of $u-v$ coverage issues with our detection. Furthermore, amplitude errors can manifest as symmetric structures, but they only follow the radio array pattern and are distinctly different from side-lobes and our detected emission pattern.

The above discussion therefore implies that even if the symmetric structure in our detection is not continuous and does not join with the core, it is still real. However, using lower \textit{gain} and number of iterations per \texttt{clean} cycle may allow us to demonstrate or verify the uniformity of emission of the dusty disk. We used a lower gain value of 0.05, which implies that at every step of a minor \texttt{clean} cycle, 5\% of the core flux is subtracted from the entire image. Reducing the number of minor cycles per iteration therefore allows more control on the imaging procedure.

\textcolor{black}{
Figure \ref{fig:alma2} shows the core as well as the core-subtracted \textit{model residuals} of four of the higher frequency ALMA bands using a lower gain. Note that these are residuals, and hence are not byproducts of any ``user-dependent" prescription. In all the images, we see a continuous diffuse structure across the location of the core. This demonstrates the uniformity of the disk emission. However, in the 654 GHz image, a ``gap" between two symmetric structures may be due to a difference in intrinsic fluxes between the two halves of the dusty disk. A slight over-subtraction of the core was needed to ``decouple" it from the disk emission, but that resulted in over-subtraction of the half with a slightly lower flux than the other. In spite of the gap, the core is very close to the brighter half of the disk. Hence this is expected even from a continuous emission. The uncertainties in fluxes are large ($\gtrsim30\%$, Table \ref{table:alma}) and can also be affected by amplitude flux calibration errors $\lesssim20\%$ at the higher frequencies. Hence these differences may be artificial, which can only be constrained by deeper spectroscopic follow-up. Even with a large $\sim20\%$ error, the $T^4$ temperature dependence results in little effect on $T$. Furthermore, a gap in the disk emission, if it really exists, would increase $r_{in}$ to $\sim0.5R$ at most. This has no effect on the outcome of the paper, since photons from the inner disk are strongly Doppler de-boosted compared to photons from the outer disk, in the co-moving frame of the jet. This has been mentioned in Section \ref{sec:dust} and Appendix \ref{sec:app2}. Another important consideration in this respect is the significant scatter in the values of the ALMA jet+dust fluxes as in Figure \ref{fig:aplib_sed2}. If the dust emission is indeed continuous, the peak flux may be contaminated with flux from the dust emission. Therefore if one simply subtracts the peak flux from the total for obtaining the jet+dust flux, considerable scatter may be expected since the radio core is variable. The only way to remedy all of the above problems perfectly is using $u-v$ plane fitting, which likely requires deeper imaging and a specialized analysis, which is out of the scope of this paper.}
%inside a modified version of \texttt{DIFMAP} (\citealt{shep94}; \texttt{ngDIFMAP}, Roychowdhury et al. in prep.) that can handle artificial effects due to variety of errors. Indeed, such a procedure has been used for the case of a faint radio-quiet AGN described in Laha et al. (in prep.). However, it is out of the scope of this paper.}

\textcolor{black}{
Given the above, a disk is the simplest model we can choose, and sub-mm photons with energy density anything similar to Equation \ref{a3} or Figure \ref{fig:uext} can produce the TeV photons, whether or not they are coming from an actual disk.}

\section{Calculation of the energy density of dust photons in the co-moving frame of the extended jet}
\label{sec:app2}

\begin{figure*}[h!]
\centering
\includegraphics[width=5in]{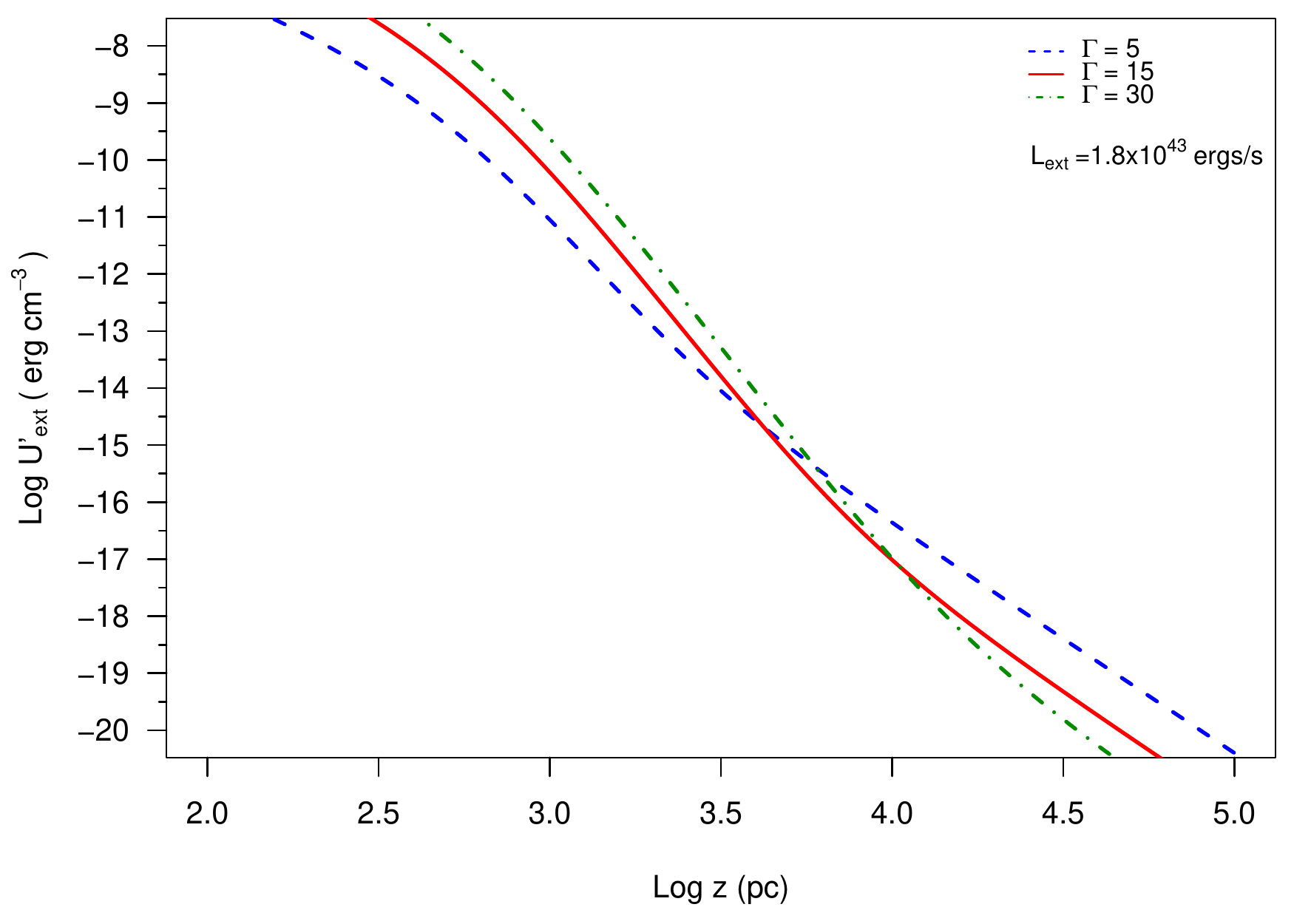}
\caption{Figure shows $U'_{ext}(z)$ for different values of $\Gamma$ as a function of de-projected distance along the jet from the jet apex for $L_{ext}=10^{43}$ ergs$/$s. As expected, higher values of $\Gamma$ provide greater beaming at small scales as well as greater de-beaming at larger scales.}
\label{fig:uext}
\end{figure*}

Assuming the dusty disk is stationary, we intend to transform the radiation energy density in the frame of the disk to the co-moving jet frame, which is moving at a bulk speed $\beta$ with respect to the disk. Any variable with primed coordinates shall refer to that in the co-moving jet frame. Therefore the external photon energy density in the jet frame (primed variables) can be written as (see e.g., \citealt{stawarz03}):

\begin{equation}
U'_{ext}=\frac{1}{c}\int_{\Omega'} I'_{ext}d\Omega '=\frac{1}{c}\int_{\Omega} I_{ext}\delta^{4}d\Omega\delta^{-2}=\frac{1}{c}\int_{\Omega} I_{ext}\delta^{2}d\Omega
\end{equation}

where $I_{ext}$ is the intensity of the seed photon field at distance $z$ along the jet axis ($I_{ext}=\delta^{4}F'_{\nu}/\Omega'$) and $\delta=[\Gamma(1+\beta\cos(\theta)]$ where $\theta$ is the angle measured from the jet axis (which is the $z$ axis) to the photon direction,  \textit{measured in the frame of the dusty disk}. $\Omega$ refers to the solid angle in the same frame. Note that we have used the fact that $\delta$ can be written equivalently as $\delta=[\Gamma(1+\beta\cos(\theta)]=[\Gamma(1-\beta\cos(\theta')]^{-1}$. We assume $I_{ext}$ is isotropic in the frame of the dusty disk, arising from a spherically symmetric emission region, which is a simplistic approximation. Therefore $I_{ext}=L_{ext}/4\pi z^2$.

Using the above equation for the dusty disk and a point on the jet axis at distance $z$ from the jet apex, it is straightforward to show that:

\begin{equation}
U'_{ext}=\frac{I_{ext}}{c}\int^{\pi}_{\pi-\tan^{-1}(R/z)} \Gamma^2[1+\beta\cos(\theta)]^2\sin(\theta)d\theta\int^{2\pi}_{0}d\phi
\end{equation}

where $R$ is the radial extent of the dusty structure defined in the main text. Note we cannot assume $\tan^{-1}(R/z)\simeq R/z\ll1$ since both the disk and the jet extend through kpc-scales. Hence we require a rigorous treatment. The complete integral, when expanded without approximations, reads:

\begin{equation}
\label{a3}
U'_{ext}(z)=\frac{L_{ext}\Gamma^2}{2 cz^2} \Bigg[(1-\mu)-\beta(1-\mu^2)+\frac{\beta^2}{3}(1-\mu^3)\Bigg]
\end{equation}

where $\mu=1/\sqrt{R^2/z^2+1}$. If we, however, use the small angle approximation by ignoring terms $\mathcal{O}(R^3/z^3)$, we obtain an expected result where the radiation is de-boosted in the co-moving jet frame:

\begin{equation}
U'_{ext}(z)\bigg\rvert_{R/z\ll1}\simeq\frac{L_{ext}R^2}{4 cz^4(1+\beta)^2\Gamma^2}
\end{equation}

In contrast, in the limit of large $R/z$, the expression assumes simple beaming, as intuitively expected. Using $R=0.65$ kpc, we plot equation \ref{a3} in Figure \ref{fig:uext} for different values of $\Gamma$ versus $z$, the distance along the jet from the jet apex. {\color{black} It is interesting to note that if the inner radius of the disk is chosen much larger than assumed, for example $r_{in}\simeq0.5R$ and if that is included in the above calculation, the resulting energy density is still unchanged since the major contribution is from the $>200$ pc part of the disk.}
%We note that the above is also consistent with results obtained in \cite{ghis96} for accretion disk photons illuminating the jet, for example.

%\clearpage

A major assumption in the above derivation is that of uniform emissivity across the disk in the $x/y$ or the polar radial direction. An approximation is provided by the GALFIT model in Table \ref{table:galfit}, where the fitted extent of the disk $R$ in the \texttt{expdisk} model essentially represents the e-folding scale for the emissivity $\simeq \exp(-r/R)$ where $r$ may represent the distance from the disk centre to a shell at a larger distance along the radial direction. Emissivity-weighted seed photon energy density will enhance the brightness of dust regions \texttt{near} the jet while reduce that for regions far away from the jet. However, photons from dust regions closer to the jet are more likely to be Dopper de-boosted and vice versa, implying the emissivity weighting will hardly change the energy density we have obtained. Hence we have omitted this for simplicity.

%\begin{equation}
%U'_{ext}(z)\simeq\frac{L_{ext}R}{3 cz^3\Gamma^2}
%\end{equation}

%\clearpage

\bibliography{version_arxiv}{}

\begin{thebibliography}{}
\expandafter\ifx\csname natexlab\endcsname\relax\def\natexlab#1{#1}\fi
\providecommand{\url}[1]{\href{#1}{#1}}
\providecommand{\dodoi}[1]{doi:~\href{http://doi.org/#1}{\nolinkurl{#1}}}
\providecommand{\doeprint}[1]{\href{http://ascl.net/#1}{\nolinkurl{http://ascl.net/#1}}}
\providecommand{\doarXiv}[1]{\href{https://arxiv.org/abs/#1}{\nolinkurl{https://arxiv.org/abs/#1}}}

\bibitem[{{Acciari} {et~al.}(2009){Acciari}, {Aliu}, {Aune}, {Beilicke},
  {Benbow}, {B{\"o}ttcher}, {Boltuch}, {Buckley}, {Bradbury}, {Bugaev},
  {Byrum}, {Cannon}, {Cesarini}, {Ciupik}, {Cogan}, {Cui}, {Dickherber},
  {Duke}, {Falcone}, {Finley}, {Fortin}, {Fortson}, {Furniss}, {Galante},
  {Gall}, {Gibbs}, {Gillanders}, {Grube}, {Guenette}, {Gyuk}, {Hanna},
  {Holder}, {Hui}, {Humensky}, {Kaaret}, {Karlsson}, {Kertzman}, {Kieda},
  {Konopelko}, {Krawczynski}, {Krennrich}, {Lang}, {Le Bohec}, {Maier},
  {McArthur}, {McCann}, {McCutcheon}, {Millis}, {Moriarty}, {Ong}, {Otte},
  {Pandel}, {Perkins}, {Pichel}, {Pohl}, {Quinn}, {Ragan}, {Reyes}, {Reynolds},
  {Roache}, {Rose}, {Sembroski}, {Smith}, {Steele}, {Theiling}, {Thibadeau},
  {Varlotta}, {Vassiliev}, {Vincent}, {Wakely}, {Ward}, {Weekes}, {Weinstein},
  {Weisgarber}, {Williams}, {Wissel}, {Wood}, {Pian}, {Vercellone},
  {Donnarumma}, {D'Ammando}, {Bulgarelli}, {Chen}, {Giuliani}, {Longo},
  {Pacciani}, {Pucella}, {Vittorini}, {Tavani}, {Argan}, {Barbiellini},
  {Caraveo}, {Cattaneo}, {Cocco}, {Costa}, {Del Monte}, {De Paris}, {Di Cocco},
  {Evangelista}, {Feroci}, {Fiorini}, {Froysland}, {Frutti}, {Fuschino},
  {Galli}, {Gianotti}, {Labanti}, {Lapshov}, {Lazzarotto}, {Lipari},
  {Marisaldi}, {Mastropietro}, {Mereghetti}, {Morelli}, {Morselli},
  {Pellizzoni}, {Perotti}, {Piano}, {Picozza}, {Pilia}, {Porrovecchio},
  {Prest}, {Rapisarda}, {Rappoldi}, {Rubini}, {Sabatini}, {Soffitta},
  {Trifoglio}, {Trois}, {Vallazza}, {Zambra}, {Zanello}, {Pittori},
  {Santolamazza}, {Verrecchia}, {Giommi}, {Colafrancesco}, {Salotti},
  {Villata}, {Raiteri}, {Aller}, {Aller}, {Arkharov}, {Efimova}, {Larionov},
  {Leto}, {Ligustri}, {Lindfors}, {Pasanen}, {Kurtanidze}, {Tetradze},
  {Lahteenmaki}, {Kotiranta}, {Cucchiara}, {Romano}, {Nesci}, {Pursimo},
  {Heidt}, {Benitez}, {Hiriart}, {Nilsson}, {Berdyugin}, {Mujica}, {Dultzin},
  {Lopez}, {Mommert}, {Sorcia}, \& {de la Calle Perez}}]{acciari09}
{Acciari}, V.~A., {Aliu}, E., {Aune}, T., {et~al.} 2009, \apj, 707, 612,
  \dodoi{10.1088/0004-637X/707/1/612}

\bibitem[{{Aharonian}(2000)}]{ahar00}
{Aharonian}, F.~A. 2000, \na, 5, 377, \dodoi{10.1016/S1384-1076(00)00039-7}

\bibitem[{{Albert} {et~al.}(2007){Albert}, {Aliu}, {Anderhub}, {Antoranz},
  {Armada}, {Baixeras}, {Barrio}, {Bartko}, {Bastieri}, {Becker}, {Bednarek},
  {Berger}, {Bigongiari}, {Biland}, {Bock}, {Bordas}, {Bosch-Ramon}, {Bretz},
  {Britvitch}, {Camara}, {Carmona}, {Chilingarian}, {Coarasa}, {Commichau},
  {Contreras}, {Cortina}, {Costado}, {Curtef}, {Danielyan}, {Dazzi}, {De
  Angelis}, {Delgado}, {de los Reyes}, {De Lotto}, {Domingo-Santamar{\'\i}a},
  {Dorner}, {Doro}, {Errando}, {Fagiolini}, {Ferenc}, {Fern{\'a}ndez}, {Firpo},
  {Flix}, {Fonseca}, {Font}, {Fuchs}, {Galante}, {Garc{\'\i}a-L{\'o}pez},
  {Garczarczyk}, {Gaug}, {Giller}, {Goebel}, {Hakobyan}, {Hayashida},
  {Hengstebeck}, {Herrero}, {H{\"o}hne}, {Hose}, {Hsu}, {Jacon}, {Jogler},
  {Kosyra}, {Kranich}, {Kritzer}, {Laille}, {Lindfors}, {Lombardi}, {Longo},
  {L{\'o}pez}, {L{\'o}pez}, {Lorenz}, {Majumdar}, {Maneva}, {Mannheim},
  {Mansutti}, {Mariotti}, {Mart{\'\i}nez}, {Mazin}, {Merck}, {Meucci}, {Meyer},
  {Miranda}, {Mirzoyan}, {Mizobuchi}, {Moralejo}, {Nilsson}, {Ninkovic},
  {O{\~n}a-Wilhelmi}, {Otte}, {Oya}, {Paneque}, {Panniello}, {Paoletti},
  {Paredes}, {Pasanen}, {Pascoli}, {Pauss}, {Pegna}, {Persic}, {Peruzzo},
  {Piccioli}, {Poller}, {Prandini}, {Puchades}, {Raymers}, {Rhode}, {Rib{\'o}},
  {Rico}, {Rissi}, {Robert}, {R{\"u}gamer}, {Saggion}, {S{\'a}nchez},
  {Sartori}, {Scalzotto}, {Scapin}, {Schmitt}, {Schweizer}, {Shayduk},
  {Shinozaki}, {Shore}, {Sidro}, {Sillanp{\"a}{\"a}}, {Sobczynska}, {Stamerra},
  {Stark}, {Takalo}, {Temnikov}, {Tescaro}, {Teshima}, {Tonello}, {Torres},
  {Turini}, {Vankov}, {Vitale}, {Wagner}, {Wibig}, {Wittek}, {Zandanel},
  {Zanin}, \& {Zapatero}}]{albert07}
{Albert}, J., {Aliu}, E., {Anderhub}, H., {et~al.} 2007, \apjl, 666, L17,
  \dodoi{10.1086/521550}

\bibitem[{{Aleksi{\'c}} {et~al.}(2014){Aleksi{\'c}}, {Ansoldi}, {Antonelli},
  {Antoranz}, {Babic}, {Bangale}, {Barres de Almeida}, {Barrio}, {Becerra
  Gonz{\'a}lez}, {Bednarek}, {Berger}, {Bernardini}, {Biland}, {Blanch},
  {Bock}, {Bonnefoy}, {Bonnoli}, {Borracci}, {Bretz}, {Carmona}, {Carosi},
  {Carreto Fidalgo}, {Colin}, {Colombo}, {Contreras}, {Cortina}, {Covino}, {Da
  Vela}, {Dazzi}, {De Angelis}, {De Caneva}, {De Lotto}, {Delgado Mendez},
  {Doert}, {Dom{\'\i}nguez}, {Dominis Prester}, {Dorner}, {Doro}, {Einecke},
  {Eisenacher}, {Elsaesser}, {Farina}, {Ferenc}, {Fonseca}, {Font}, {Frantzen},
  {Fruck}, {Garc{\'\i}a L{\'o}pez}, {Garczarczyk}, {Garrido Terrats}, {Gaug},
  {Giavitto}, {Godinovi{\'c}}, {Gonz{\'a}lez Mu{\~n}oz}, {Gozzini}, {Hadamek},
  {Hadasch}, {Herrero}, {Hildebrand}, {Hose}, {Hrupec}, {Idec}, {Kadenius},
  {Kellermann}, {Knoetig}, {Krause}, {Kushida}, {La Barbera}, {Lelas},
  {Lewandowska}, {Lindfors}, {Lombardi}, {L{\'o}pez}, {L{\'o}pez-Coto},
  {L{\'o}pez-Oramas}, {Lorenz}, {Lozano}, {Makariev}, {Mallot}, {Maneva},
  {Mankuzhiyil}, {Mannheim}, {Maraschi}, {Marcote}, {Mariotti},
  {Mart{\'\i}nez}, {Mazin}, {Menzel}, {Meucci}, {Miranda}, {Mirzoyan},
  {Moralejo}, {Munar-Adrover}, {Nakajima}, {Niedzwiecki}, {Nilsson}, {Nowak},
  {Orito}, {Overkemping}, {Paiano}, {Palatiello}, {Paneque}, {Paoletti},
  {Paredes}, {Paredes-Fortuny}, {Partini}, {Persic}, {Prada}, {Prada Moroni},
  {Prandini}, {Preziuso}, {Puljak}, {Reinthal}, {Rhode}, {Rib{\'o}}, {Rico},
  {Rodriguez Garcia}, {R{\"u}gamer}, {Saggion}, {Saito}, {Saito}, {Salvati},
  {Satalecka}, {Scalzotto}, {Scapin}, {Schultz}, {Schweizer}, {Shore},
  {Sillanp{\"a}{\"a}}, {Sitarek}, {Snidaric}, {Sobczynska}, {Spanier},
  {Stamatescu}, {Stamerra}, {Steinbring}, {Storz}, {Sun}, {Suri{\'c}},
  {Takalo}, {Tavecchio}, {Terzi{\'c}}, {Tescaro}, {Teshima}, {Thaele},
  {Tibolla}, {Torres}, {Toyama}, {Treves}, {Uellenbeck}, {Vogler}, {Wagner},
  {Zandanel}, {Zanin}, {MAGIC Collaboration}, {Balmaverde}, {Kataoka},
  {Rekola}, \& {Takahashi}}]{ngc1275}
{Aleksi{\'c}}, J., {Ansoldi}, S., {Antonelli}, L.~A., {et~al.} 2014, \aap, 564,
  A5, \dodoi{10.1051/0004-6361/201322951}

\bibitem[{{Alexander} \& {Hickox}(2012)}]{alex12}
{Alexander}, D.~M., \& {Hickox}, R.~C. 2012, \nar, 56, 93,
  \dodoi{10.1016/j.newar.2011.11.003}

\bibitem[{{Anderhub} {et~al.}(2009){Anderhub}, {Antonelli}, {Antoranz},
  {Backes}, {Baixeras}, {Balestra}, {Barrio}, {Bastieri}, {Becerra
  Gonz{\'a}lez}, {Becker}, {Bednarek}, {Berdyugin}, {Berger}, {Bernardini},
  {Biland}, {Bock}, {Bonnoli}, {Bordas}, {Borla Tridon}, {Bosch-Ramon}, {Bose},
  {Braun}, {Bretz}, {Britzger}, {Camara}, {Carmona}, {Carosi}, {Colin},
  {Commichau}, {Contreras}, {Cortina}, {Costado}, {Covino}, {Dazzi}, {De
  Angelis}, {de Cea del Pozo}, {De los Reyes}, {De Lotto}, {De Maria}, {De
  Sabata}, {Delgado Mendez}, {Dom{\'\i}nguez}, {Dominis Prester}, {Dorner},
  {Doro}, {Elsaesser}, {Errando}, {Ferenc}, {Fern{\'a}ndez}, {Firpo},
  {Fonseca}, {Font}, {Galante}, {Garc{\'\i}a L{\'o}pez}, {Garczarczyk}, {Gaug},
  {Godinovic}, {Goebel}, {Hadasch}, {Herrero}, {Hildebrand},
  {H{\"o}hne-M{\"o}nch}, {Hose}, {Hrupec}, {Hsu}, {Jogler}, {Klepser},
  {Kranich}, {La Barbera}, {Laille}, {Leonardo}, {Lindfors}, {Lombardi},
  {Longo}, {L{\'o}pez}, {Lorenz}, {Majumdar}, {Maneva}, {Mankuzhiyil},
  {Mannheim}, {Maraschi}, {Mariotti}, {Mart{\'\i}nez}, {Mazin}, {Meucci},
  {Miranda}, {Mirzoyan}, {Miyamoto}, {Mold{\'o}n}, {Moles}, {Moralejo},
  {Nieto}, {Nilsson}, {Ninkovic}, {Orito}, {Oya}, {Paoletti}, {Paredes},
  {Pasanen}, {Pascoli}, {Pauss}, {Pegna}, {Perez-Torres}, {Persic}, {Peruzzo},
  {Prada}, {Prandini}, {Puchades}, {Puljak}, {Reichardt}, {Rhode}, {Rib{\'o}},
  {Rico}, {Rissi}, {Robert}, {R{\"u}gamer}, {Saggion}, {Sainio}, {Saito},
  {Salvati}, {S{\'a}nchez-Conde}, {Satalecka}, {Scalzotto}, {Scapin},
  {Schweizer}, {Shayduk}, {Shore}, {Sierpowska-Bartosik}, {Sillanp{\"a}{\"a}},
  {Sitarek}, {Sobczynska}, {Spanier}, {Spiro}, {Stamerra}, {Stark}, {Suric},
  {Takalo}, {Tavecchio}, {Temnikov}, {Tescaro}, {Teshima}, {Torres}, {Turini},
  {Vankov}, {Wagner}, {Villforth}, {Zabalza}, {Zandanel}, {Zanin}, \&
  {Zapatero}}]{anderhub09}
{Anderhub}, H., {Antonelli}, L.~A., {Antoranz}, P., {et~al.} 2009, \apjl, 704,
  L129, \dodoi{10.1088/0004-637X/704/2/L129}

\bibitem[{{Asada} {et~al.}(2009){Asada}, {Kameno}, {Shen}, {Shinji}, {Gabuzda},
  \& {Inoue}}]{asad09}
{Asada}, K., {Kameno}, S., {Shen}, Z.-Q., {et~al.} 2009, in Astronomical
  Society of the Pacific Conference Series, Vol. 402, Approaching
  Micro-Arcsecond Resolution with VSOP-2: Astrophysics and Technologies, ed.
  Y.~{Hagiwara}, E.~{Fomalont}, M.~{Tsuboi}, \& M.~{Yasuhiro}, 91

\bibitem[{{Baxter} {et~al.}(1987){Baxter}, {Disney}, \& {Phillipps}}]{baxter87}
{Baxter}, D.~A., {Disney}, M.~J., \& {Phillipps}, S. 1987, \mnras, 228, 313,
  \dodoi{10.1093/mnras/228.2.313}

\bibitem[{{Blandford} {et~al.}(2019){Blandford}, {Meier}, \&
  {Readhead}}]{blandford2019}
{Blandford}, R., {Meier}, D., \& {Readhead}, A. 2019, \araa, 57, 467,
  \dodoi{10.1146/annurev-astro-081817-051948}

\bibitem[{{Blandford} \& {Payne}(1982)}]{bland82}
{Blandford}, R.~D., \& {Payne}, D.~G. 1982, \mnras, 199, 883,
  \dodoi{10.1093/mnras/199.4.883}

\bibitem[{{Blandford} \& {Znajek}(1977)}]{bland77}
{Blandford}, R.~D., \& {Znajek}, R.~L. 1977, \mnras, 179, 433,
  \dodoi{10.1093/mnras/179.3.433}

\bibitem[{{B{\"o}ttcher}(2007)}]{bott07}
{B{\"o}ttcher}, M. 2007, \apss, 309, 95, \dodoi{10.1007/s10509-007-9404-0}

\bibitem[{{B{\"o}ttcher} \& {Els}(2016)}]{bottcher16}
{B{\"o}ttcher}, M., \& {Els}, P. 2016, \apj, 821, 102,
  \dodoi{10.3847/0004-637X/821/2/102}

\bibitem[{{Breiding} {et~al.}(2017){Breiding}, {Meyer}, {Georganopoulos},
  {Keenan}, {DeNigris}, \& {Hewitt}}]{breiding17}
{Breiding}, P., {Meyer}, E.~T., {Georganopoulos}, M., {et~al.} 2017, \apj, 849,
  95, \dodoi{10.3847/1538-4357/aa907a}

\bibitem[{{Briggs}(1995)}]{briggs95}
{Briggs}, D.~S. 1995, in American Astronomical Society Meeting Abstracts, Vol.
  187, American Astronomical Society Meeting Abstracts, 112.02

\bibitem[{{Cassaro} {et~al.}(1999){Cassaro}, {Stanghellini}, {Bondi},
  {Dallacasa}, {della Ceca}, \& {Zappal{\`a}}}]{cas99}
{Cassaro}, P., {Stanghellini}, C., {Bondi}, M., {et~al.} 1999, \aaps, 139, 601,
  \dodoi{10.1051/aas:1999511}

\bibitem[{{Castignani} \& {De Zotti}(2015)}]{castig15}
{Castignani}, G., \& {De Zotti}, G. 2015, \aap, 573, A125,
  \dodoi{10.1051/0004-6361/201423903}

\bibitem[{{Chiaberge} \& {Marconi}(2011)}]{chiaberge11}
{Chiaberge}, M., \& {Marconi}, A. 2011, \mnras, 416, 917,
  \dodoi{10.1111/j.1365-2966.2011.19079.x}

\bibitem[{{de Jong} {et~al.}(2015){de Jong}, {Beckmann}, {Soldi}, {Tramacere},
  \& {Gros}}]{m87}
{de Jong}, S., {Beckmann}, V., {Soldi}, S., {Tramacere}, A., \& {Gros}, A.
  2015, \mnras, 450, 4333, \dodoi{10.1093/mnras/stv927}

\bibitem[{{de Koff} {et~al.}(2000){de Koff}, {Best}, {Baum}, {Sparks},
  {R{\"o}ttgering}, {Miley}, {Golombek}, {Macchetto}, \& {Martel}}]{dekoff00}
{de Koff}, S., {Best}, P., {Baum}, S.~A., {et~al.} 2000, \apjs, 129, 33,
  \dodoi{10.1086/313402}

\bibitem[{{Disney} {et~al.}(1974){Disney}, {Peterson}, \& {Rodgers}}]{disney74}
{Disney}, M.~J., {Peterson}, B.~A., \& {Rodgers}, A.~W. 1974, \apjl, 194, L79,
  \dodoi{10.1086/181673}

\bibitem[{{Drouart} {et~al.}(2012){Drouart}, {De Breuck}, {Vernet}, {Laing},
  {Seymour}, {Stern}, {Haas}, {Pier}, \& {Rocca-Volmerange}}]{drouart12}
{Drouart}, G., {De Breuck}, C., {Vernet}, J., {et~al.} 2012, \aap, 548, A45,
  \dodoi{10.1051/0004-6361/201220059}

\bibitem[{{Ellison} {et~al.}(2011){Ellison}, {Patton}, {Mendel}, \&
  {Scudder}}]{ellison11}
{Ellison}, S.~L., {Patton}, D.~R., {Mendel}, J.~T., \& {Scudder}, J.~M. 2011,
  \mnras, 418, 2043, \dodoi{10.1111/j.1365-2966.2011.19624.x}

\bibitem[{{Fossati} {et~al.}(1998){Fossati}, {Maraschi}, {Celotti}, {Comastri},
  \& {Ghisellini}}]{fos98}
{Fossati}, G., {Maraschi}, L., {Celotti}, A., {Comastri}, A., \& {Ghisellini},
  G. 1998, \mnras, 299, 433, \dodoi{10.1046/j.1365-8711.1998.01828.x}

\bibitem[{{Fritz} {et~al.}(2006){Fritz}, {Franceschini}, \&
  {Hatziminaoglou}}]{fritz06}
{Fritz}, J., {Franceschini}, A., \& {Hatziminaoglou}, E. 2006, \mnras, 366,
  767, \dodoi{10.1111/j.1365-2966.2006.09866.x}

\bibitem[{{Gebhardt} {et~al.}(2000){Gebhardt}, {Bender}, {Bower}, {Dressler},
  {Faber}, {Filippenko}, {Green}, {Grillmair}, {Ho}, {Kormendy}, {Lauer},
  {Magorrian}, {Pinkney}, {Richstone}, \& {Tremaine}}]{geb00}
{Gebhardt}, K., {Bender}, R., {Bower}, G., {et~al.} 2000, \apjl, 539, L13,
  \dodoi{10.1086/312840}

\bibitem[{{Georganopoulos} {et~al.}(2005){Georganopoulos}, {Kazanas},
  {Perlman}, \& {Stecker}}]{georg05}
{Georganopoulos}, M., {Kazanas}, D., {Perlman}, E., \& {Stecker}, F.~W. 2005,
  \apj, 625, 656, \dodoi{10.1086/429558}

\bibitem[{{Georganopoulos} {et~al.}(2006){Georganopoulos}, {Perlman},
  {Kazanas}, \& {McEnery}}]{georg06}
{Georganopoulos}, M., {Perlman}, E.~S., {Kazanas}, D., \& {McEnery}, J. 2006,
  \apjl, 653, L5, \dodoi{10.1086/510452}

\bibitem[{{Ghisellini} {et~al.}(1993){Ghisellini}, {Padovani}, {Celotti}, \&
  {Maraschi}}]{ghis93}
{Ghisellini}, G., {Padovani}, P., {Celotti}, A., \& {Maraschi}, L. 1993, \apj,
  407, 65, \dodoi{10.1086/172493}

\bibitem[{{Ghisellini} \& {Tavecchio}(2010)}]{ghis10}
{Ghisellini}, G., \& {Tavecchio}, F. 2010, \mnras, 409, L79,
  \dodoi{10.1111/j.1745-3933.2010.00952.x}

\bibitem[{{Hardcastle} \& {Croston}(2020)}]{hardcastle20}
{Hardcastle}, M.~J., \& {Croston}, J.~H. 2020, \nar, 88, 101539,
  \dodoi{10.1016/j.newar.2020.101539}

\bibitem[{{Harris} \& {Krawczynski}(2006)}]{harris06}
{Harris}, D.~E., \& {Krawczynski}, H. 2006, \araa, 44, 463,
  \dodoi{10.1146/annurev.astro.44.051905.092446}

\bibitem[{{Harvey} {et~al.}(2020){Harvey}, {Georganopoulos}, \&
  {Meyer}}]{harvey20}
{Harvey}, A. L.~W., {Georganopoulos}, M., \& {Meyer}, E.~T. 2020, Nature
  Communications, 11, 5475, \dodoi{10.1038/s41467-020-19296-6}

\bibitem[{{Hervet} {et~al.}(2015){Hervet}, {Boisson}, \& {Sol}}]{hervet2015}
{Hervet}, O., {Boisson}, C., \& {Sol}, H. 2015, \aap, 578, A69,
  \dodoi{10.1051/0004-6361/201425330}

\bibitem[{{HESS Collaboration} {et~al.}(2015){HESS Collaboration},
  {Abramowski}, {Aharonian}, {Ait Benkhali}, {Akhperjanian}, {Ang{\"u}ner},
  {Anton}, {Backes}, {Balenderan}, {Balzer}, {Barnacka}, {Becherini}, {Becker
  Tjus}, {Bernl{\"o}hr}, {Birsin}, {Bissaldi}, {Biteau}, {B{\"o}ttcher},
  {Boisson}, {Bolmont}, {Bordas}, {Brucker}, {Brun}, {Brun}, {Bulik},
  {Carrigan}, {Casanova}, {Chadwick}, {Chalme-Calvet}, {Chaves},
  {Cheesebrough}, {Chr{\'e}tien}, {Colafrancesco}, {Cologna}, {Conrad},
  {Couturier}, {Cui}, {Dalton}, {Daniel}, {Davids}, {Degrange}, {Deil},
  {deWilt}, {Dickinson}, {Djannati-Ata{\"\i}}, {Domainko}, {O'C. Drury},
  {Dubus}, {Dutson}, {Dyks}, {Dyrda}, {Edwards}, {Egberts}, {Eger}, {Espigat},
  {Farnier}, {Fegan}, {Feinstein}, {Fernand es}, {Fernandez}, {Fiasson},
  {Fontaine}, {F{\"o}rster}, {F{\"u}{\ss}ling}, {Gajdus}, {Gallant},
  {Garrigoux}, {Giavitto}, {Giebels}, {Glicenstein}, {Grondin},
  {Grudzi{\'n}ska}, {H{\"a}ffner}, {Hahn}, {Harris}, {Heinzelmann}, {Henri},
  {Hermann}, {Hervet}, {Hillert}, {Hinton}, {Hofmann}, {Hofverberg}, {Holler},
  {Horns}, {Jacholkowska}, {Jahn}, {Jamrozy}, {Janiak}, {Jankowsky}, {Jung},
  {Kastendieck}, {Katarzy{\'n}ski}, {Katz}, {Kaufmann}, {Kh{\'e}lifi},
  {Kieffer}, {Klepser}, {Klochkov}, {Klu{\'z}niak}, {Kneiske}, {Kolitzus},
  {Komin}, {Kosack}, {Krakau}, {Krayzel}, {Kr{\"u}ger}, {Laffon}, {Lamanna},
  {Lefaucheur}, {Lemi{\`e}re}, {Lemoine-Goumard}, {Lenain}, {Lohse}, {Lopatin},
  {Lu}, {Marandon}, {Marcowith}, {Marx}, {Maurin}, {Maxted}, {Mayer}, {McComb},
  {M{\'e}hault}, {Meintjes}, {Menzler}, {Meyer}, {Moderski}, {Mohamed},
  {Moulin}, {Murach}, {Naumann}, {de Naurois}, {Niemiec}, {Nolan}, {Oakes},
  {Odaka}, {Ohm}, {de O{\~n}a Wilhelmi}, {Opitz}, {Ostrowski}, {Oya}, {Panter},
  {Parsons}, {Paz Arribas}, {Pekeur}, {Pelletier}, {Perez}, {Petrucci},
  {Peyaud}, {Pita}, {Poon}, {P{\"u}hlhofer}, {Punch}, {Quirrenbach}, {Raab},
  {Raue}, {Reichardt}, {Reimer}, {Reimer}, {Renaud}, {de los Reyes}, {Rieger},
  {Rob}, {Romoli}, {Rosier-Lees}, {Rowell}, {Rudak}, {Rulten}, {Sahakian},
  {Sanchez}, {Santangelo}, {Schlickeiser}, {Sch{\"u}ssler}, {Schulz},
  {Schwanke}, {Schwarzburg}, {Schwemmer}, {Sol}, {Spengler}, {Spies},
  {Stawarz}, {Steenkamp}, {Stegmann}, {Stinzing}, {Stycz}, {Sushch},
  {Tavernet}, {Tavernier}, {Taylor}, {Terrier}, {Tluczykont}, {Trichard},
  {Valerius}, {van Eldik}, {van Soelen}, {Vasileiadis}, {Venter}, {Viana},
  {Vincent}, {V{\"o}lk}, {Volpe}, {Vorster}, {Vuillaume}, {Wagner}, {Wagner},
  {Wagner}, {Ward}, {Weidinger}, {Weitzel}, {White}, {Wierzcholska},
  {Willmann}, {W{\"o}rnlein}, {Wouters}, {Yang}, {Zabalza}, {Zacharias},
  {Zdziarski}, {Zech}, {Zechlin}, {Finke}, {Fortin}, \& {Horan}}]{hess15}
{HESS Collaboration}, {Abramowski}, A., {Aharonian}, F., {et~al.} 2015, \aap,
  573, A31, \dodoi{10.1051/0004-6361/201321436}

\bibitem[{{Hickox} \& {Alexander}(2018)}]{hickox18}
{Hickox}, R.~C., \& {Alexander}, D.~M. 2018, \araa, 56, 625,
  \dodoi{10.1146/annurev-astro-081817-051803}

\bibitem[{{H{\"o}nig}(2019)}]{honig19}
{H{\"o}nig}, S.~F. 2019, \apj, 884, 171, \dodoi{10.3847/1538-4357/ab4591}

\bibitem[{{Hopkins} \& {Quataert}(2010)}]{hopkinsquat10}
{Hopkins}, P.~F., \& {Quataert}, E. 2010, \mnras, 407, 1529,
  \dodoi{10.1111/j.1365-2966.2010.17064.x}

\bibitem[{{Hovatta} {et~al.}(2009){Hovatta}, {Valtaoja}, {Tornikoski}, \&
  {L{\"a}hteenm{\"a}ki}}]{hovatta09}
{Hovatta}, T., {Valtaoja}, E., {Tornikoski}, M., \& {L{\"a}hteenm{\"a}ki}, A.
  2009, \aap, 494, 527, \dodoi{10.1051/0004-6361:200811150}

\bibitem[{{Jones} {et~al.}(2009){Jones}, {Read}, {Saunders}, {Colless},
  {Jarrett}, {Parker}, {Fairall}, {Mauch}, {Sadler}, {Watson}, {Burton},
  {Campbell}, {Cass}, {Croom}, {Dawe}, {Fiegert}, {Frankcombe}, {Hartley},
  {Huchra}, {James}, {Kirby}, {Lahav}, {Lucey}, {Mamon}, {Moore}, {Peterson},
  {Prior}, {Proust}, {Russell}, {Safouris}, {Wakamatsu}, {Westra}, \&
  {Williams}}]{jones09}
{Jones}, D.~H., {Read}, M.~A., {Saunders}, W., {et~al.} 2009, \mnras, 399, 683,
  \dodoi{10.1111/j.1365-2966.2009.15338.x}

\bibitem[{{Joshi} \& {B{\"o}ttcher}(2007)}]{joshi07}
{Joshi}, M., \& {B{\"o}ttcher}, M. 2007, \apj, 662, 884, \dodoi{10.1086/518210}

\bibitem[{{Kaufmann} {et~al.}(2013){Kaufmann}, {Wagner}, \& {Tibolla}}]{kauf13}
{Kaufmann}, S., {Wagner}, S.~J., \& {Tibolla}, O. 2013, \apj, 776, 68,
  \dodoi{10.1088/0004-637X/776/2/68}

\bibitem[{{Keenan} {et~al.}(2021){Keenan}, {Meyer}, {Georganopoulos}, {Reddy},
  \& {French}}]{keen21}
{Keenan}, M., {Meyer}, E.~T., {Georganopoulos}, M., {Reddy}, K., \& {French},
  O.~J. 2021, \mnras, 505, 4726, \dodoi{10.1093/mnras/stab1182}

\bibitem[{{Lambrides} {et~al.}(2021){Lambrides}, {Chiaberge}, {Heckman},
  {Kirkpatrick}, {Meyer}, {Petric}, {Hall}, {Long}, {Watts}, {Gilli}, {Simons},
  {Tchernyshyov}, {Rodriguez-Gomez}, {Vito}, {De La Vega}, {Davis}, {Kocevski},
  \& {Norman}}]{lambr21}
{Lambrides}, E., {Chiaberge}, M., {Heckman}, T., {et~al.} 2021, arXiv e-prints,
  arXiv:2107.07533.
\newblock \doarXiv{2107.07533}

\bibitem[{{Landt} {et~al.}(2010){Landt}, {Buchanan}, \& {Barmby}}]{landt10}
{Landt}, H., {Buchanan}, C.~L., \& {Barmby}, P. 2010, \mnras, 408, 1982,
  \dodoi{10.1111/j.1365-2966.2010.17264.x}

\bibitem[{{Lister} {et~al.}(2019){Lister}, {Homan}, {Hovatta}, {Kellermann},
  {Kiehlmann}, {Kovalev}, {Max-Moerbeck}, {Pushkarev}, {Readhead}, {Ros}, \&
  {Savolainen}}]{lister19}
{Lister}, M.~L., {Homan}, D.~C., {Hovatta}, T., {et~al.} 2019, \apj, 874, 43,
  \dodoi{10.3847/1538-4357/ab08ee}

\bibitem[{{Madejski} \& {Sikora}(2016)}]{mades16}
{Madejski}, G.~G., \& {Sikora}, M. 2016, \araa, 54, 725,
  \dodoi{10.1146/annurev-astro-081913-040044}

\bibitem[{{Madejski} {et~al.}(2016){Madejski}, {Nalewajko}, {Madsen}, {Chiang},
  {Balokovi{\'c}}, {Paneque}, {Furniss}, {Hayashida}, {Urry}, {Sikora},
  {Ajello}, {Blandford}, {Harrison}, {Sanchez}, {Giebels}, {Stern},
  {Alexander}, {Barret}, {Boggs}, {Christensen}, {Craig}, {Forster}, {Giommi},
  {Grefenstette}, {Hailey}, {Hornstrup}, {Kitaguchi}, {Koglin}, {Mao},
  {Miyasaka}, {Mori}, {Perri}, {Pivovaroff}, {Puccetti}, {Rana}, {Westergaard},
  {Zhang}, \& {Zoglauer}}]{madej16}
{Madejski}, G.~M., {Nalewajko}, K., {Madsen}, K.~K., {et~al.} 2016, \apj, 831,
  142, \dodoi{10.3847/0004-637X/831/2/142}

\bibitem[{{Mannheim}(1993)}]{mann93}
{Mannheim}, K. 1993, \prd, 48, 2408, \dodoi{10.1103/PhysRevD.48.2408}

\bibitem[{{Mannheim} {et~al.}(1991){Mannheim}, {Biermann}, \&
  {Kruells}}]{mann91}
{Mannheim}, K., {Biermann}, P.~L., \& {Kruells}, W.~M. 1991, \aap, 251, 723

\bibitem[{{McMullin} {et~al.}(2007){McMullin}, {Waters}, {Schiebel}, {Young},
  \& {Golap}}]{casa}
{McMullin}, J.~P., {Waters}, B., {Schiebel}, D., {Young}, W., \& {Golap}, K.
  2007, in Astronomical Society of the Pacific Conference Series, Vol. 376,
  Astronomical Data Analysis Software and Systems XVI, ed. R.~A. {Shaw},
  F.~{Hill}, \& D.~J. {Bell}, 127

\bibitem[{{Mehta} {et~al.}(2009){Mehta}, {Georganopoulos}, {Perlman},
  {Padgett}, \& {Chartas}}]{mehta09}
{Mehta}, K.~T., {Georganopoulos}, M., {Perlman}, E.~S., {Padgett}, C.~A., \&
  {Chartas}, G. 2009, \apj, 690, 1706, \dodoi{10.1088/0004-637X/690/2/1706}

\bibitem[{{Meyer} {et~al.}(2011){Meyer}, {Fossati}, {Georganopoulos}, \&
  {Lister}}]{meyer11}
{Meyer}, E.~T., {Fossati}, G., {Georganopoulos}, M., \& {Lister}, M.~L. 2011,
  \apj, 740, 98, \dodoi{10.1088/0004-637X/740/2/98}

\bibitem[{{Meyer} {et~al.}(2012){Meyer}, {Fossati}, {Georganopoulos}, \&
  {Lister}}]{meyer12}
---. 2012, \apjl, 752, L4, \dodoi{10.1088/2041-8205/752/1/L4}

\bibitem[{{Meyer} \& {Georganopoulos}(2014)}]{meyer14}
{Meyer}, E.~T., \& {Georganopoulos}, M. 2014, \apjl, 780, L27,
  \dodoi{10.1088/2041-8205/780/2/L27}

\bibitem[{{Meyer} {et~al.}(2015){Meyer}, {Georganopoulos}, {Sparks}, {Godfrey},
  {Lovell}, \& {Perlman}}]{meyer15}
{Meyer}, E.~T., {Georganopoulos}, M., {Sparks}, W.~B., {et~al.} 2015, \apj,
  805, 154, \dodoi{10.1088/0004-637X/805/2/154}

\bibitem[{{M{\"u}cke} \& {Protheroe}(2001)}]{muecke01}
{M{\"u}cke}, A., \& {Protheroe}, R.~J. 2001, Astroparticle Physics, 15, 121,
  \dodoi{10.1016/S0927-6505(00)00141-9}

\bibitem[{{M{\"u}cke} {et~al.}(2003){M{\"u}cke}, {Protheroe}, {Engel},
  {Rachen}, \& {Stanev}}]{mucke03}
{M{\"u}cke}, A., {Protheroe}, R.~J., {Engel}, R., {Rachen}, J.~P., \& {Stanev},
  T. 2003, Astroparticle Physics, 18, 593,
  \dodoi{10.1016/S0927-6505(02)00185-8}

\bibitem[{{Nenkova} {et~al.}(2002){Nenkova}, {Ivezi{\'c}}, \&
  {Elitzur}}]{nenk02}
{Nenkova}, M., {Ivezi{\'c}}, {\v{Z}}., \& {Elitzur}, M. 2002, \apjl, 570, L9,
  \dodoi{10.1086/340857}

\bibitem[{{Peng} {et~al.}(2002){Peng}, {Ho}, {Impey}, \& {Rix}}]{peng02}
{Peng}, C.~Y., {Ho}, L.~C., {Impey}, C.~D., \& {Rix}, H.-W. 2002, \aj, 124,
  266, \dodoi{10.1086/340952}

\bibitem[{{Perucho} {et~al.}(2014){Perucho}, {Mart{\'\i}}, {Laing}, \&
  {Hardee}}]{perucho14}
{Perucho}, M., {Mart{\'\i}}, J.~M., {Laing}, R.~A., \& {Hardee}, P.~E. 2014,
  \mnras, 441, 1488, \dodoi{10.1093/mnras/stu676}

\bibitem[{{Pesce} {et~al.}(1994){Pesce}, {Falomo}, \& {Treves}}]{pesce94}
{Pesce}, J.~E., {Falomo}, R., \& {Treves}, A. 1994, \aj, 107, 494,
  \dodoi{10.1086/116871}

\bibitem[{{Peterson} {et~al.}(1976){Peterson}, {Rodgers}, {Wampler}, \&
  {Disney}}]{peterson76}
{Peterson}, B.~A., {Rodgers}, A.~W., {Wampler}, E.~J., \& {Disney}, M.~J. 1976,
  \apjl, 207, L17, \dodoi{10.1086/182168}

\bibitem[{{Petropoulou} \& {Dimitrakoudis}(2015)}]{petrop15_2}
{Petropoulou}, M., \& {Dimitrakoudis}, S. 2015, \mnras, 452, 1303,
  \dodoi{10.1093/mnras/stv1380}

\bibitem[{{Petropoulou} {et~al.}(2015){Petropoulou}, {Dimitrakoudis},
  {Padovani}, {Mastichiadis}, \& {Resconi}}]{petrop15}
{Petropoulou}, M., {Dimitrakoudis}, S., {Padovani}, P., {Mastichiadis}, A., \&
  {Resconi}, E. 2015, \mnras, 448, 2412, \dodoi{10.1093/mnras/stv179}

\bibitem[{{Petropoulou} {et~al.}(2014){Petropoulou}, {Lefa}, {Dimitrakoudis},
  \& {Mastichiadis}}]{petrop14}
{Petropoulou}, M., {Lefa}, E., {Dimitrakoudis}, S., \& {Mastichiadis}, A. 2014,
  \aap, 562, A12, \dodoi{10.1051/0004-6361/201322833}

\bibitem[{{Petropoulou} {et~al.}(2017){Petropoulou}, {Vasilopoulos}, \&
  {Giannios}}]{petr17}
{Petropoulou}, M., {Vasilopoulos}, G., \& {Giannios}, D. 2017, \mnras, 464,
  2213, \dodoi{10.1093/mnras/stw2453}

\bibitem[{{Petry} \& {CASA Development Team}(2012)}]{petry12}
{Petry}, D., \& {CASA Development Team}. 2012, in Astronomical Society of the
  Pacific Conference Series, Vol. 461, Astronomical Data Analysis Software and
  Systems XXI, ed. P.~{Ballester}, D.~{Egret}, \& N.~P.~F. {Lorente}, 849.
\newblock \doarXiv{1201.3454}

\bibitem[{{Plotkin} {et~al.}(2012){Plotkin}, {Anderson}, {Brandt}, {Markoff},
  {Shemmer}, \& {Wu}}]{plot12}
{Plotkin}, R.~M., {Anderson}, S.~F., {Brandt}, W.~N., {et~al.} 2012, \apjl,
  745, L27, \dodoi{10.1088/2041-8205/745/2/L27}

\bibitem[{{Rodgers} \& {Peterson}(1977)}]{rodgers77}
{Rodgers}, A.~W., \& {Peterson}, B.~A. 1977, \apjl, 212, L9,
  \dodoi{10.1086/182363}

\bibitem[{{Roustazadeh} \& {B{\"o}ttcher}(2011)}]{rous11}
{Roustazadeh}, P., \& {B{\"o}ttcher}, M. 2011, \apj, 728, 134,
  \dodoi{10.1088/0004-637X/728/2/134}

\bibitem[{{Sanchez} {et~al.}(2015){Sanchez}, {Giebels}, {Fortin}, {Horan},
  {Szostek}, {Fegan}, {Baczko}, {Finke}, {Kadler}, {Kovalev}, {Lister},
  {Pushkarev}, \& {Savolainen}}]{san15}
{Sanchez}, D.~A., {Giebels}, B., {Fortin}, P., {et~al.} 2015, \mnras, 454,
  3229, \dodoi{10.1093/mnras/stv2151}

\bibitem[{{Schlafly} \& {Finkbeiner}(2011)}]{schlafly11}
{Schlafly}, E.~F., \& {Finkbeiner}, D.~P. 2011, \apj, 737, 103,
  \dodoi{10.1088/0004-637X/737/2/103}

\bibitem[{{Sikora}(2011)}]{sikora11}
{Sikora}, M. 2011, in Jets at All Scales, ed. G.~E. {Romero}, R.~A. {Sunyaev},
  \& T.~{Belloni}, Vol. 275, 59--67, \dodoi{10.1017/S1743921310015644}

\bibitem[{{Sikora} {et~al.}(1996){Sikora}, {Sol}, {Begelman}, \&
  {Madejski}}]{sikora96}
{Sikora}, M., {Sol}, H., {Begelman}, M.~C., \& {Madejski}, G.~M. 1996, \mnras,
  280, 781, \dodoi{10.1093/mnras/280.3.781}

\bibitem[{{Sikora} {et~al.}(2009){Sikora}, {Stawarz}, {Moderski}, {Nalewajko},
  \& {Madejski}}]{sikora09}
{Sikora}, M., {Stawarz}, {\L}., {Moderski}, R., {Nalewajko}, K., \& {Madejski},
  G.~M. 2009, \apj, 704, 38, \dodoi{10.1088/0004-637X/704/1/38}

\bibitem[{{Smithsonian Astrophysical Observatory}(2000)}]{ds9}
{Smithsonian Astrophysical Observatory}. 2000, {SAOImage DS9: A utility for
  displaying astronomical images in the X11 window environment}.
\newblock \doeprint{0003.002}

\bibitem[{{Sparks} {et~al.}(2000){Sparks}, {Baum}, {Biretta}, {Macchetto}, \&
  {Martel}}]{sparks00}
{Sparks}, W.~B., {Baum}, S.~A., {Biretta}, J., {Macchetto}, F.~D., \& {Martel},
  A.~R. 2000, \apj, 542, 667, \dodoi{10.1086/317064}

\bibitem[{{Stalevski} {et~al.}(2012){Stalevski}, {Fritz}, {Baes}, {Nakos}, \&
  {Popovi{\'c}}}]{stalev12}
{Stalevski}, M., {Fritz}, J., {Baes}, M., {Nakos}, T., \& {Popovi{\'c}},
  L.~{\v{C}}. 2012, \mnras, 420, 2756, \dodoi{10.1111/j.1365-2966.2011.19775.x}

\bibitem[{{Stawarz} {et~al.}(2003){Stawarz}, {Sikora}, \&
  {Ostrowski}}]{stawarz03}
{Stawarz}, {\L}., {Sikora}, M., \& {Ostrowski}, M. 2003, \apj, 597, 186,
  \dodoi{10.1086/378290}

\bibitem[{{Stickel} {et~al.}(1993){Stickel}, {Fried}, \& {Kuehr}}]{stickel94}
{Stickel}, M., {Fried}, J.~W., \& {Kuehr}, H. 1993, \aaps, 98, 393

\bibitem[{{Tadhunter}(2016)}]{tadh16}
{Tadhunter}, C. 2016, \aapr, 24, 10, \dodoi{10.1007/s00159-016-0094-x}

\bibitem[{{Tanada} {et~al.}(2019){Tanada}, {Kataoka}, \& {Inoue}}]{tanada19}
{Tanada}, K., {Kataoka}, J., \& {Inoue}, Y. 2019, \apj, 878, 139,
  \dodoi{10.3847/1538-4357/ab2233}

\bibitem[{{Tavecchio} {et~al.}(2010){Tavecchio}, {Ghisellini}, {Ghirlanda},
  {Foschini}, \& {Maraschi}}]{tav10}
{Tavecchio}, F., {Ghisellini}, G., {Ghirlanda}, G., {Foschini}, L., \&
  {Maraschi}, L. 2010, \mnras, 401, 1570,
  \dodoi{10.1111/j.1365-2966.2009.15784.x}

\bibitem[{{Tavecchio} {et~al.}(2000){Tavecchio}, {Maraschi}, {Sambruna}, \&
  {Urry}}]{tav00}
{Tavecchio}, F., {Maraschi}, L., {Sambruna}, R.~M., \& {Urry}, C.~M. 2000,
  \apjl, 544, L23, \dodoi{10.1086/317292}

\bibitem[{{The GIMP Development Team}(2019)}]{gimp}
{The GIMP Development Team}. 2019, GIMP, 2.10.12.
\newblock \url{https://www.gimp.org}

\bibitem[{{Urry} \& {Mushotzky}(1982)}]{urry82}
{Urry}, C.~M., \& {Mushotzky}, R.~F. 1982, \apj, 253, 38,
  \dodoi{10.1086/159607}

\bibitem[{{Urry} \& {Padovani}(1995)}]{urry95}
{Urry}, C.~M., \& {Padovani}, P. 1995, \pasp, 107, 803, \dodoi{10.1086/133630}

\bibitem[{{von Montigny} {et~al.}(1995){von Montigny}, {Bertsch}, {Chiang},
  {Dingus}, {Esposito}, {Fichtel}, {Fierro}, {Hartman}, {Hunter}, {Kanbach},
  {Kniffen}, {Lin}, {Mattox}, {Mayer-Hasselwander}, {Michelson}, {Nolan},
  {Radecke}, {Schneid}, {Sreekumar}, {Thompson}, \& {Willis}}]{vonm95}
{von Montigny}, C., {Bertsch}, D.~L., {Chiang}, J., {et~al.} 1995, \apj, 440,
  525, \dodoi{10.1086/175294}

\bibitem[{{Wakely} \& {Horan}(2008)}]{tevcat08}
{Wakely}, S.~P., \& {Horan}, D. 2008, International Cosmic Ray Conference, 3,
  1341

\bibitem[{{Woo} {et~al.}(2005){Woo}, {Urry}, {van der Marel}, {Lira}, \&
  {Maza}}]{woo05}
{Woo}, J.-H., {Urry}, C.~M., {van der Marel}, R.~P., {Lira}, P., \& {Maza}, J.
  2005, \apj, 631, 762, \dodoi{10.1086/432681}

\bibitem[{{Zacharias} \& {Wagner}(2016)}]{zac16}
{Zacharias}, M., \& {Wagner}, S. 2016, Galaxies, 4, 63,
  \dodoi{10.3390/galaxies4040063}

\bibitem[{{Zdziarski} \& {Bottcher}(2015)}]{zdz15}
{Zdziarski}, A.~A., \& {Bottcher}, M. 2015, \mnras, 450, L21,
  \dodoi{10.1093/mnrasl/slv039}

\end{thebibliography}
\bibliographystyle{aasjournal}

\end{document}